\documentclass{aastex631}

\usepackage{graphicx}
\usepackage{amsmath}
\usepackage{xcolor}
\usepackage{relsize}
\usepackage{xspace}

\newcommand{\todo}[1]{}
\newcommand{\textbftog}[1]{#1}

\DeclareMathOperator\erf{erf}

\def\CTA{Center for Theoretical Astrophysics, Los Alamos National Laboratory, Los Alamos, NM 87545, USA}
\def\CCS{Computer, Computational, and Statistical Sciences Division, Los Alamos National Laboratory, Los Alamos, NM
 87545, USA}
\def\CNLS{Center for Nonlinear Studies, Los Alamos National Laboratory, Los Alamos, NM
 87545, USA}
\def\T5{T-5 Applied Mathematics and Plasma Physics Group, Los Alamos National Laboratory, Los Alamos, NM 87545, USA}
\def\XCP{Computational Physics Division, Los Alamos National Laboratory, Los Alamos, NM, 87545, USA}
\def\UA{The University of Arizona, Tucson, AZ 85721, USA}
\def\NM{Department of Physics and Astronomy, The University of New Mexico, Albuquerque, NM 87131, USA}
\def\GWU{The George Washington University, Washington, DC 20052, USA}

\graphicspath{{./}{figures/}}

\begin{document}

\title{On a spectral method for $\beta$-particle bound excitation collisions in kilonovae}

\author[0000-0003-3265-4079]{Ryan~T. Wollaeger}
\affiliation{\CTA}
\affiliation{\CCS}

\author[0000-0003-2624-0056]{Chris~L. Fryer}
\affiliation{\CNLS}
\affiliation{\CTA}
\affiliation{\UA}
\affiliation{\NM}
\affiliation{\GWU}

\author[0000-0003-4582-9894]{Robert Chiodi}
\affiliation{\CCS}

\author[0000-0002-4906-2195]{Peter T. Brady}
\affiliation{\CCS}

\author[0000-0003-4156-5342]{Oleg Korobkin}
\affiliation{\CTA}
\affiliation{\T5}

\author[0000-0002-2447-3131]{Cale Harnish}
\affiliation{\CCS}

\author[0000-0003-1087-2964]{Christopher~J. Fontes}
\affiliation{\CTA}
\affiliation{\XCP}

\author[0000-0002-2044-0885]{Jeffrey~R. Haack}
\affiliation{\CCS}

\author[0000-0002-5321-2838]{Oleksandr Chapurin}
\affiliation{\T5}

\author[0000-0002-0935-760X]{Oleksandr Koshkarov}
\affiliation{\T5}

\author[0000-0002-7030-2683]{Gian Luca Delzanno}
\affiliation{\T5}

\author[0000-0003-2367-1547]{Daniel Livescu}
\affiliation{\CCS}



\begin{abstract}

The interaction of $\beta$-particles with the weakly ionized plasma background is an
important mechanism for powering the kilonova transient signal from neutron star mergers.
For this purpose, we present an implementation of the approximate fast-particle collision
kernel, described by \cite{inokuti1971} following the seminal formulation of \cite{bethe1930}, in a spectral
solver of the Vlasov-Maxwell-Boltzmann equations.
In particular, we expand the fast-particle plane-wave atomic excitation kernel into
coefficients of the Hermite basis, and derive the relevant discrete spectral system.
In this fast-particle limit, the approach permits the direct use of atomic data, including
optical oscillator strengths, normally applied to photon-matter interaction.
The resulting spectral matrix is implemented in the MASS-APP spectral solver framework,
in a way that avoids full matrix storage per spatial zone.
We numerically verify aspects of the matrix construction, and present a proof-of-principle 3D
simulation of a 2D axisymmetric kilonova ejecta snapshot.
Our preliminary numerical results indicate that a reasonable choice of Hermite basis parameters for
$\beta$-particles in the kilonova are a bulk velocity parameter $\vec{u}=0$, a thermal velocity
parameter $\vec{\alpha}=0.5c$, and a 9x9x9 mode velocity basis set (Hermite orders 0 to 8
in each dimension).
\textbftog{For ejecta-interior sample zones, we estimate the ratio of thermalization from
large-angle ($\gtrsim2.5^{\circ}$) bound excitation scattering to total thermalization
is $\sim$0.002-0.003.}

\end{abstract}



\section{Introduction} \label{sec:intro}

Kilonovae (KNe) are radioactively powered electromagnetic (EM) transients signaling
the aftermath of double neutron star or neutron star-black hole binary mergers
(an incomplete sequence of KN model developments up to 2017 might be given by
\cite{lattimer1974,lattimer1976,li1998,freiburghaus1999,roberts2011,kasen2013,tanaka2013,fonte2015,barnes2016,metzger2017}).
As the two compact objects inspiral due to the emission of gravitational waves, the
neutron star(s) will be tidally disrupted, causing neutron-rich mass to eject and become
gravitationally unbound.
The merger results in a compact remnant (neutron star or black hole) surrounded by an
accretion disk, from which various mechanisms produce further (``post-merger'') ejecta
(see, for example,~\cite{perego2014,martin2015,desai2022}).
The detailed EM spectra and broadband magnitudes from the observation of KN AT2017gfo
(see, for example,
\cite{arcavi2017,cowperthwaite2017,drout2017,kasliwal2017,smartt2017,tanvir2017,troja2017,villar2017}),
in concert with the gravitational
wave observation GW170817~\citep{abbott2017a,abbott2017b}, provided an unprecedented window
into the pre- and post-merger phases of the transient, and by examining the nuclear decay
pattern in the ejecta seemed to confirm neutron star mergers are a source of r-process
elements~\citep{rosswog2018}.

While the basic picture of KNe has remained unchanged for several decades, since the
semi-analytic work of~\cite{li1998} (for a review of KN physics, see also~\textbftog{\cite{metzger2019}}),
simulating KNe at high fidelity is an ever-developing field.
Recent studies explore detailed radiative transfer, atomic physics, non-local thermodynamic
equilibrium (non-LTE), and multidimensional spatial ejecta, for instance:
\cite{fontes2020,tanaka2020,fontes2023} on detailed LTE opacity,
\cite{hotokezaka2021,pognan2022a,pognan2022b,pognan2023} on detailed non-LTE opacity,
and~\cite{korobkin2021,heinzel2021,bulla2023,fryer2023} on spatial distribution and
multidimensional spatial effects (and see references therein).
Each of these aspects brings a level of uncertainty into the simulations, that otherwise
might be encapsulated in free parameters (for example, grey opacity).
A large source of uncertainty in state-of-the-art calculations, particularly at late times
as the KN ejecta becomes nebular, is in modeling the interaction of decay products
($\alpha$, $\beta$, and $\gamma$ particles) with the ions forming the ejecta.
In full generality, it is a complicated problem of multiple particle fields undergoing
transfer and interaction with atomic orbital structure and free electrons.
Moreover, the atomic structure of the elements formed in r-process can involve tens of millions
of resonances between thousands to millions of energy levels (see, for example, the atomic
data presented by~\cite{fontes2020,tanaka2020}).

The process of $\beta$-particle thermalization (the loss of particle kinetic energy to
Coulomb interactions, atomic excitations, ionizations, etc.) is inefficient compared
to thermalization of the more massive $\alpha$ particles or fission fragments
\citep{barnes2016,barnes2021,zhu2021}, and hence is non-local ($\beta$-particles travel significant
length scales relative to the ejecta to deposit energy).
\cite{barnes2016,barnes2021} and~\cite{zhu2021} have demonstrated this inefficiency significantly
impacts the observable EM KN signal (on the order of a factor of 2 in luminosity, for instance).
The state-of-the-art detailed thermalization model for $\beta$-particles presented by~\cite{barnes2021}
uses the Bethe stopping-potential prescription~\citep{bethe1930}, which accounts for energy loss over a particle path
due to multiple small-angle scatters and encapsulates atomic properties with an average ionization.
A question remains as to the impact of large-angle scatters that induce excitation effects in the
ion background, which in principle requires the so-called ``generalized'' oscillator strengths that were introduced by \cite{bethe1930} and
elaborated on by~\cite{inokuti1971}.

Consequently, having a framework for implementing atomic data directly into a thermalization
calculation, along with a formulation for particle transfer that can be extended to different
differential cross sections, is useful for making inroads to improved fidelity.
We attempt one such preliminary inroad using a deterministic spectral plasma solver implemented
in the CPU/GPU-parallel Multiphysics Adaptive Scalable Simulator for Applications in Plasma Physics
(MASS-APP) code base~\citep{chiodi2024}.
``Spectral'' in this context implies the particle phase space distribution function is expanded
over a complete basis function set in velocity space.
To undertake proof-of-concept simulations, we implement the non-relativistic inelastic scattering
kernel for excitation, described by~\cite{inokuti1971}. 
The inelastic form of the two-body, or binary, integral collision kernel is given by~\cite{garibotti1994},
and has been used before in the context of spectral Boltzmann methods (see, for instance the
spectral-Lagrangian Boltzmann equation solver by~\cite{munafo2014}).
Here we specifically employ an asymmetrically weighted Hermite function basis~\citep{armstrong1970},
which has a beneficial property of bridging macro (fluid)-micro (kinetic) scales (see, for example,
\cite{camporeale2006,vencels2015,delzanno2015,koshkarov2021}).
The derivational sequence we present here is similar to the Hermite expansion of the multi-species
collision kernel presented by~\cite{wang2019,li2022,li2023}, but we do not expand both distributions
into the binary kernel, instead expanding one and also expanding the cross section itself in the
Hermite basis.
We also make approximations to the kernel that make the integral over the ion species separable
from the integral over $\beta$-particle velocity (which generally will not hold outside the
fast-particle approximation).
This permits us to employ an efficient closed form for integrals of products involving three
Hermite polynomials and two Gaussian weights, which to our knowledge is not used by
\cite{wang2019,li2022,li2023}, since their kernel expansion does not isolate these terms
(to be sure, the approaches of these authors are more general, in being able to solve for multiple
distribution species).

Neglecting internal conversion, $\beta$-particle emission has a smooth continuum
\citep{fermi1934,schenter1983,alekseev2021} for a spectrum, lending itself well to smooth
basis functions.
Hence, we see this spectral Hermite basis technique as a possible deterministic supplement to
Particle-in-Cell (PIC) or Monte Carlo methods that may be better suited to treating sharp
\textbftog{electron} emission lines resulting from internal conversion.
Moreover, deterministic schemes of course do not have stochastic noise, so they may be well suited to
capturing large-angle scattering effects, specifically.

This paper is organized as follows.
In Section~\ref{sec:method}, we write the governing equations solved with MASS-APP and describe the
derivation of the excitation collision kernel used for $\beta$-particles.
In Section~\ref{sec:numver}, we present numerical verification of particular aspects of the method,
comparing closed form derivations to direct numerical integration of terms that build the collision
matrix needed for simulation.
In Section~\ref{sec:3dkn}, we describe a trial $\beta$-particle simulation of excitation interactions
for a 2D axisymmetric morphology embedded in 3D Cartesian geometry.
Finally, in Section~\ref{sec:conclude} we summarize our findings and discuss future work that would
further improve fidelity.

\section{Spectral Method and Implementation} \label{sec:method}

In this section, we write down the full spectrally discretized set of equations.
Subsequently, we focus attention on incorporating the non-relativistic differential excitation
cross section given by~\cite{inokuti1971} into the spectral basis framework, and discuss the
approximations made to simplify the collision kernel.
After deriving the spectral Hermite form of the fast-particle kernel, we provide an outline
summarizing the calculations of this section, including how the steps may be extensible to other differential
cross sections.
Supplementary detail is provided in Appendix~\ref{app:A} for the evaluation of the two-body collision
kernel and in Appendix~\ref{app:B} for the closed-form evaluation of integrals involving
three Hermite polynomials and two Gaussian weight functions, referred to henceforth as ``compact triple Hermite products'',
which are used to evaluate each term in the spectrally discrete collision matrix derived in this section.

The Vlasov-Maxwell-Boltzmann system of equations we consider is
\begin{subequations}
    \label{eq1:method}
    \begin{gather}
        \frac{\partial f_s}{\partial t} + \vec{v}\cdot\nabla f_s
        + \frac{q_s}{m_s}(\vec{E}+\vec{v}\times\vec{B})\cdot\nabla_v f_s = \mathcal{C}[f_a, f_s] \;\;, \\
        \frac{\partial\vec{E}}{\partial t} =   c\nabla\times\vec{B}  - 4 \pi \vec{J} \;\;, \\
        \frac{\partial\vec{B}}{\partial t} = - c\nabla\times \vec{E} \;\;,  \\
        \vec{J} = q_s \int \vec{v} f_s d^3 \vec{v} + q_a \int \vec{v} f_a d^3 \vec{v} \;\;,
    \end{gather}
\end{subequations}
where $t$ is time, $\vec{v}$ is velocity, the $\nabla$ operator is the gradient with
respect to the spatial coordinate ($\vec{x}$), $\nabla_v$ is the gradient operator with
respect to velocity, subscript $s$ indicates the species ($\beta$-particles here),
subscript $a$ indicates the atom/ion background, $q_s$ is charge, $m_s$ is mass, $\vec{E}$ and
$\vec{B}$ are the electric and magnetic fields, $f_s$ and $f_a$ are the $\beta$-particle and
background distributions (number density per spatial volume per velocity volume), $\vec{J}$
is the current density\textbftog{, and $\mathcal{C}[\cdot,\cdot]$ is the collision operator}.
For the purpose of this work, we set the interacting distribution of ejecta atom/ions as
given within a discrete time step, thus linearizing the collision term.
This approach is consistent with the practice of computing decay thermalization and radiative
transfer in separate steps from the ejecta plasma state update, but may incur error
where moderate bulk thermodynamic changes lead to significant electron occupation number discrepancy
for particular atomic states.

The Hermite basis is orthogonal with respect to a Gaussian weight, hence amenable to determining
expansion coefficients via inner products.
Following~\cite{delzanno2015}, we expand the distribution as
\begin{equation}
    \label{eq2:method}
    f_s(\vec{x},\vec{v},t) = \sum_{n,m,p}C_{n,m,p}(\vec{x},t)\Psi_{n,m,p}(\vec{\xi}) \;\;,
\end{equation}
where the subscripts $n$, $m$, and $p$ are the order of the basis function in each velocity
dimension, $C_{n,m,p}$ is the expansion coefficient (for which we solve), and the basis
function $\Psi_{n,m,p}(\vec{\xi})$ is given by Eq.~\eqref{eq3:appB}.
The $\vec{\xi}$ argument of the basis functions is non-dimensional velocity (subscript $xyz$ indicates
component),
\begin{equation}
    \label{eq3:method}
    \vec{\xi} = (\xi_x, \xi_y, \xi_z)
    = \left(\frac{v_x-u_x}{\alpha_x}, \frac{v_y-u_y}{\alpha_y}, \frac{v_z-u_z}{\alpha_z}\right)
    = (\vec{v} - \vec{u}) \oslash \vec{\alpha} \;\;,
\end{equation}
where $\vec{u}$ and $\vec{\alpha}$ are user-provided, constant velocity parameters corresponding to bulk
and thermal velocity in the Gaussian factor of the basis.
\textbftog{The last equality represents $\vec{\xi}$ with Hadamard (element-wise) division.}
 
The system of equations resulting from expanding Eqs.~\eqref{eq1:method}
with Eq.~\eqref{eq2:method} and taking Hermite inner products is (see, for example,~\cite{delzanno2015})
\begin{subequations}
    \label{eq4:method}
    \begin{gather}
        \frac{\partial C_{n,m,p}}{\partial t}
        + \;\mathlarger{\ldots}\; = \sum_{n',m',p'}S_{n,m,p}^{n',m',p'}C_{n',m',p'} \;\;, \\
        \frac{\partial\vec{B}}{\partial t} = -c\nabla \times \vec{E} \;\;, \\
        \frac{\partial\vec{E}}{\partial t} =  c\nabla \times \vec{B} 
        - 4 \pi q_s \alpha_x  \alpha_y  \alpha_z \left (C_{0,0,0} 
        \begin{bmatrix}
            u_x \\
            u_y \\
            u_z 
        \end{bmatrix}
        + \frac{1}{\sqrt{2}} 
        \begin{bmatrix}
            \alpha_x C_{1,0,0} \\
            \alpha_y C_{0,1,0} \\
            \alpha_z C_{0,0,1} 
        \end{bmatrix}\right) \;\;,
    \end{gather}
\end{subequations}
where $S_{n,m,p}^{n',m',p'}$ is a collision matrix dependent on the background ion properties
of the KN ejecta, and in Eq.~\eqref{eq4:method}c we have made the approximation that the ion
background does not contribute significant current density.
For brevity, on the left side of Eq.~\eqref{eq4:method}a, we have omitted a spatial divergence
operator including the flux and Lorentz force from $\vec{E}$ and $\vec{B}$, but include them in
a version of the equations in Appendix~\ref{app:C}.

\subsection{Fast-particle, non-relativistic, differential cross section}

We derive the non-relativistic form of the scattering matrix $S_{n,m,p}^{n',m',p'}$ using the
differential cross section in the Bethe-Born (high-energy) limit, presented by~\cite{inokuti1971} for excitation from
atomic state $j'$ to atomic state $j$,
\begin{equation}
    \label{eq1:cros}
    \frac{d\sigma_{jj'}}{d\Omega} = 4\left(\frac{Me^2}{\hbar^2}\right)^2
    \left(\frac{k'}{k}\right)K^{-4}\left(\frac{R}{E_{jj'}}\right)(Ka_0)^2f_{jj'}(K) \;\;,
\end{equation}
where
\begin{equation}
    \label{eq2:cros}
    M = \frac{m_eM_a}{m_e + M_a}
\end{equation}
is the reduced mass ($m_e$ and $M_a$ are electron and atomic mass), $e$ is the electron
charge, $\hbar$ is the reduced Planck constant, $k$ ($k'$) is the incoming (outgoing)
wave number (equivalently momentum $=\hbar k$) of the free electron, $K$ is the magnitude
of the difference of the incoming and outgoing wave numbers, $R = m_ee^4/2\hbar^2=13.606$ eV
is the Rydberg energy, $E_{jj'}$ is the energy difference between levels $j$ and $j'$,
$a_0 = \hbar^2/(m_ee^2)=0.52918\times10^{-8}$ cm is the Bohr radius, and $f_{jj'}(K)$ is the
generalized oscillator strength.
The magnitudes of wave number obey the law of cosines, 
\begin{equation}
    \label{eq3:cros}
    K^2 = k^2 + (k')^2 - 2k'k\cos(\theta) \;\;,
\end{equation}
where $\theta$ is the polar angle of deflection in the center-of-mass frame.
If the magnitude of pre- and post-deflection wave numbers is taken to be independent of
the polar deflection angle, then
\begin{equation}
    \label{eq5:cros}
    \frac{d(K^2)}{d\theta} = 2k'k\sin(\theta) \;\;,
\end{equation}
which is given by~\cite{inokuti1971} to replace the solid angle differential
$d\Omega$ with $d(K^2)$.
Assuming $k'=k$, then $K^2=2k^2(1-\cos(\theta))=4k^2\sin^2(\theta/2)$, and Eq.
\eqref{eq1:cros} becomes
\begin{equation}
    \label{eq5:cros}
    \frac{d\sigma_{jj'}}{d\Omega} = \left(\frac{Me^2}{\hbar^2}\right)^2
    \frac{\csc^4(\theta/2)}{4k^{4}}
    \left[\left(\frac{R}{E_{jj'}}\right)(Ka_0)^2f_{jj'}(K)\right] \;\;,
\end{equation}
Supposing an elastic collision, conservation of kinetic energy and momentum imply
\begin{equation}
    \label{eq6:cros}
    k' = k\left(\frac{\cos(\theta)+\sqrt{(M_a/m_e)^2-\sin^2(\theta)}}{1 + M_a/m_e}\right)
    \;\;,
\end{equation}
where the higher root is taken, so that the solution would be correct if $M_a=m_e$.
Consequently,
\begin{equation}
    \label{eq7:cros}
    \lim_{M_a/m_e\rightarrow\infty}k' = k \;\;,
\end{equation}
so the condition of $k'=k$, giving Eq.~\eqref{eq5:cros} is equivalent to $m_e\ll M_a$
for elastic collisions.
Also in this limit, the reduced mass converges to the electron mass $m_e$, and noting
$\hbar k = m_ev_0$, where $v_0$ is the initial electron velocity, Eq.~\eqref{eq5:cros}
becomes
\begin{equation}
    \label{eq8:cros}
    \frac{d\sigma_{jj'}}{d\Omega}
    = \left(\frac{e^2}{2m_ev_0^2}\right)^2\csc^4(\theta/2)
    \left[\left(\frac{R}{E_{jj'}}\right)(Ka_0)^2f_{jj'}(K)\right] \;\;,
\end{equation}
where the coefficient outside the brackets on the right side is now the standard
non-relativistic definition of classical Rutherford scattering
(a more general form is noted by~\cite{inokuti1971} as the coefficient, not making
the assumption of $m_e \ll M_a$).
According to~\cite{inokuti1971}, the term in the square brackets is
the conditional probability that the atom will excite from state $j'$ to state $j$,
given a magnitude of momentum exchange from the electron of $K$.

It is notable that, $k=k'$ does not imply $K$ is small; the angle $\theta$ has to
vanish for $K$ to vanish.
Using Eq.~\eqref{eq3:cros} and conservation of kinetic energy, balanced with excitation,
the formula for $(Ka_0)^2$ in terms of initial electron kinetic energy $E_k=mv_0^2/2$
and $\theta$ is~\citep{inokuti1971}
\begin{equation}
    \label{eq9:cros}
    (Ka_0)^2 = 2\frac{E_k}{R}\left(\frac{M}{m_e}\right)^2
    \left[1-\frac{1}{2}\left(\frac{m_eE_{jj'}}{ME_k}\right)
    -\left(\sqrt{1-\left(\frac{m_eE_{jj'}}{ME_k}\right)}\right)\cos(\theta)
    \right] \;\;.
\end{equation}
$(Ka_0)^2$ has the following limits, corresponding to $\theta=0$ or $\pi$ in
Eq.~\eqref{eq9:cros}~\citep{inokuti1971}:
\begin{subequations}
    \label{eq10:cros}
    \begin{gather}
        (Ka_0)_{\min}^2 \approx \frac{1}{4}\left(\frac{E_{jj'}^2}{RE_k}\right)
        \left[1+\frac{1}{2}\left(\frac{m_eE_{jj'}}{MR}\right)\right] \;\;,\\
        (Ka_0)_{\max}^2 \approx 4\frac{E_k}{R}\left(\frac{M}{m_e}\right)^2
        \left[1-\frac{1}{2}\left(\frac{m_eE_{jj'}}{ME_k}\right)\right] \;\;,
    \end{gather}
\end{subequations}
where the square root coefficient of $\cos(\theta)$ has been Taylor expanded
in $m_eE_{jj'}/(ME_k)$~\citep{inokuti1971}.
The assumption that $E_k \gg E_{jj'}$ and $E_k \gg R$ implies the maximum value of
$(Ka_0)_{\max}^2 \gg 1$.
However, the Rutherford coefficient of Eq.~\eqref{eq8:cros} grows rapidly as $\theta$ vanishes.
Moreover, the generalized oscillator strength decreases rapidly for large $Ka_0$
(see~\cite{inokuti1971}, Section 3.2).
Thus, low $Ka_0$ and small-angle deflections, i.e. forwarding scattering, are most probable from the Bethe kernel.
If the generalized oscillator strength is expanded in a Taylor series as a function of $Ka_0$~\citep{inokuti1971}, and
assuming $(Ka_0)^2\sim (Ka_0)_{\min}^2$ due to the dominance of forward scattering, it becomes reasonable to replace the generalized
oscillator strength with the optical oscillator strength, $f_{jj'} \approx f_{jj'}(K=0)$, which is the essence of the Bethe high-energy approximation.

If we take $f_{jj'} = f_{jj'}(K=0)$, we may readily use the oscillator strength data
used for photon opacities in Eq.~\eqref{eq8:cros}.
Doing so, and also incorporating Eq.~\eqref{eq9:cros} into Eq.~\eqref{eq8:cros}, we have
a non-relativistic excitation cross section from state $j'$ to state $j$,
\begin{equation}
    \label{eq11:cros}
    \frac{d\sigma_{jj'}}{d\Omega}
    = 2\left(\frac{e^2}{4E_k}\right)^2\csc^4(\theta/2)
    \left(\frac{E_k}{E_{jj'}}\right)f_{jj'}
    \left[1-\frac{1}{2}\left(\frac{E_{jj'}}{E_k}\right)
    -\left(\sqrt{1-\left(\frac{E_{jj'}}{E_k}\right)}\right)\cos(\theta)
    \right]
    \;\;,
\end{equation}
where $m_e=M$ has been incorporated into Eq.~\eqref{eq9:cros}.
The only quantity with units on the right side is $(e^2/4E_k)^2$ which should have units
of length-squared.
In electrostatic cgs units (used by Inokuti 1971), so
\begin{equation}
    \label{eq12:cros}
    \left(\frac{e^2}{4E_k}\right)^2 
    \approx \left(\frac{197}{548}\frac{1}{E_{k,{\rm MeV}}}\right)^2 {\rm fm}^2 \;\;,
\end{equation}
where $E_{k,{\rm MeV}}$ is the initial kinetic energy in MeV and fm=femtometers.

\subsection{Spectral matrix from binary collision kernel}

We now incorporate Eq.~\eqref{eq11:cros} into the two-body inelastic collision kernel
(the elastic version is given by~\cite{mihalas1984}, Chapter 1, for example).
For our purpose, we decompose the atomic distribution into sub-distributions corresponding
to each excited state $j'$, and expand the collision kernel as follows,
\begin{equation}
    \label{eq2:awhb}
    \mathcal{C}[f_a,f_s] = \sum_{(j,j')\in\mathcal{T}}\mathcal{C}_{jj'}[f_{a,j'},f_{a,j},f_s] \;\;,
\end{equation}
where $\mathcal{T}$ is the set of possible state transition pairs $(j,j')$,
\begin{equation}
    \label{eq3:awhb}
    \mathcal{C}_{jj'}[f_{a,j'},f_{a,j},f_s] = \int\int|\vec{v}_a-\vec{v}_s|\frac{d\sigma_{jj'}}{d\Omega}
    \left(\frac{g_{j'}}{g_j}f_s(\vec{v}_s')f_{a,j}(\vec{v}_a')
    -f_s(\vec{v}_s)f_{a,j'}(\vec{v}_a)\right)d\Omega d^3\vec{v}_a \;\;,
\end{equation}
and we have subscripted velocities to indicate ion (``a'') or $\beta$-particle (``s'').
We note this expression follows~\cite{munafo2014}, where the energy level degeneracy
is included as a coefficient of the product of distributions over post-scattered velocities.
The distribution prime superscript indicates evaluation at the post-collision momentum.

Post-scatter velocities $\vec{v}_a'$ and $\vec{v}_s'$ can be evaluated from conservation of
momentum and energy in the non-relativistic limit, giving
\begin{subequations}
    \label{eq5:awhb}
    \begin{gather}
        \vec{v}_s' = \frac{1}{m_e+M_a}
        \left(m_e\vec{v}_s + M_a\vec{v}_a
        + M_a\left(\sqrt{v_{sa}^2-\frac{2E_{jj'}}{M}}\right)\hat{\Omega}'\right)
        \;\;, \\
        \vec{v}_a' = \frac{1}{m_e+M_a}
        \left(m_e\vec{v}_s + M_a\vec{v}_a
        - m_e\left(\sqrt{v_{sa}^2-\frac{2E_{jj'}}{M}}\right)\hat{\Omega}'\right)
        \;\;,
    \end{gather}
\end{subequations}
where $v_{sa}=|\vec{v}_{sa}|=|\vec{v}_a-\vec{v}_s|$.
Equations~\eqref{eq5:awhb} show the post-scatter velocities as a function of pre-scatter
velocity and energy weight, which must be incorporated into Eq.~\eqref{eq3:awhb}
to evaluate the post-scatter portion of the integral~\citep{garibotti1994}.

Considering only excitation collisions in Eq.~\eqref{eq1:method}, and expanding the
$\beta$-distribution $f_s$ with Eq.~\eqref{eq3:method},
\begin{multline}
    \label{eq5:method}
    \sum_{n,m,p}\frac{\partial C_{n,m,p}}{\partial t}\Psi_{n,m,p}(\vec{\xi})
    = \sum_{n,m,p}C_{n,m,p}\int\int v_{sa}\frac{d\sigma_{jj'}}{d\Omega}
    \frac{g_{j'}}{g_j}\Psi_{n,m,p}(\vec{\xi}')f_{a,j}(\vec{v}_a')d\Omega d^3\vec{v}_a \\
    - \sum_{n,m,p}C_{n,m,p}\int\int v_{sa}\frac{d\sigma_{jj'}}{d\Omega}
    \Psi_{n,m,p}(\vec{\xi})f_{a,j'}(\vec{v}_a)d\Omega d^3\vec{v}_a \;\;,
\end{multline}
where $\vec{\xi}'$ is the non-dimensional form of the post-scatter velocity $\vec{v}_s'$.
Taking the basis inner product, Eq.~\eqref{eq5:method} becomes
\begin{multline}
    \label{eq6:method}
    \frac{\partial C_{n',m',p'}}{\partial t} =
    \sum_{n,m,p}C_{n,m,p}\int\Psi^{n',m',p'}(\vec{\xi})\int\int v_{sa}\frac{d\sigma_{jj'}}{d\Omega}
    \frac{g_{j'}}{g_j}\Psi_{n,m,p}(\vec{\xi}')f_{a,j}(\vec{v}_a')d\Omega d^3\vec{v}_a d^3\vec{\xi} \\
    - \sum_{n,m,p}C_{n,m,p}\int\Psi^{n',m',p'}(\vec{\xi})\int\int v_{sa}\frac{d\sigma_{jj'}}{d\Omega}
    \Psi_{n,m,p}(\vec{\xi})f_{a,j'}(\vec{v}_a)d\Omega d^3\vec{v}_a d^3\vec{\xi} \;\;.
\end{multline}
The right side of Eq.~\eqref{eq6:method} is now in the form of a matrix product with the
spectral solution vector $C_{n,m,p}$, as in the right side of Eq.~\eqref{eq4:method}.
Following~\cite{munafo2014}, the integral in the first summation on the right side of~\eqref{eq6:method}
permits us to use the principle of micro-reversibility,
\begin{equation}
    \label{eq7:method}
    v_{sa}\frac{d\sigma_{jj'}}{d\Omega}d\Omega d^3\vec{v}_a d^3\vec{\xi}
    = \frac{g_j}{g_{j'}}
    v_{sa}'\frac{d\sigma_{j'j}}{d\Omega'}d\Omega' d^3\vec{v}_a' d^3\vec{\xi}'
    \;\;,
\end{equation}
where we have taken $d\Omega'$ to be the differential solid angle about the pre-scatter direction $\hat{\Omega}$,
to simplify Eq.~\eqref{eq6:method} to
\begin{multline}
    \label{eq8:method}
    \frac{\partial C_{n',m',p'}}{\partial t} =
    \sum_{n,m,p}C_{n,m,p}\left[\int\int v_{sa}'\left(\int\Psi^{n',m',p'}(\vec{\xi}(\vec{\xi}',\vec{v}_a',\vec{\Omega}))
    \frac{d\sigma_{j'j}}{d\Omega'}d\Omega'\right)
    \Psi_{n,m,p}(\vec{\xi}')f_{a,j}(\vec{v}_a') d^3\vec{v}_a' d^3\vec{\xi}' \right.\\\left.
    - \int\Psi^{n',m',p'}(\vec{\xi})\int v_{sa}\left(\int\frac{d\sigma_{jj'}}{d\Omega}d\Omega\right)
    \Psi_{n,m,p}(\vec{\xi})f_{a,j'}(\vec{v}_a) d^3\vec{v}_a d^3\vec{\xi}\right] \;\;.
\end{multline}
However, the upper-indexed basis is still a function of pre-scatter velocity $\vec{\xi}$, which is
now a function of the post-scatter velocities and collision angle (time-inverting Eq.~\eqref{eq5:awhb}).

In order to further simplify Eq.~\eqref{eq8:method}, we make some approximations that should be reasonable,
at least for the fast-particle approximation and the conditions of the proper inertial frame of the KN ejecta.
The first approximation is to evaluate $\Psi^{n',m',p'}(\vec{\xi}(\vec{\xi}',\vec{v}_a',\vec{\Omega}))$ at
assumed average values of 0 for the pre-scatter angle $\vec{\Omega}$ and the post-scatter ion velocity $\vec{v}_a$,
thus permitting factoring out $\Psi^{n',m',p'}$ from the first integral, giving
\begin{multline}
    \label{eq9:method}
    \frac{\partial C_{n',m',p'}}{\partial t} =
    \sum_{n,m,p}C_{n,m,p}\left[\int\int v_{sa}'\Psi^{n',m',p'}
    \left(\frac{1}{m_e + M_a}\left(m_e\vec{\xi}'-M_a\vec{u}\oslash\vec{\alpha}\right)\right)
    \sigma_{j'j}\Psi_{n,m,p}(\vec{\xi}')f_{a,j}(\vec{v}_a') d^3\vec{v}_a' d^3\vec{\xi}'
    \right.\\\left.
    - \int\Psi^{n',m',p'}(\vec{\xi})\int v_{sa}\sigma_{jj'}
    \Psi_{n,m,p}(\vec{\xi})f_{a,j'}(\vec{v}_a) d^3\vec{v}_a d^3\vec{\xi}\right] \;\;.
\end{multline}
where use has been made of Eq.~\eqref{eq5:awhb}a with the substitution of the 0-averages to obtain
the new argument of $\Psi^{n',m',p'}$ in the first integral, and \textbftog{$\vec{u}\oslash\vec{\alpha}$ 
again denotes the element-wise division of $\vec{u}$ by $\vec{\alpha}$ (Hadamard division).}
The isolated differential cross sections in the first and second integral have been integrated over pre- and
post-scattering solid angle, respectively, giving $\sigma_{j',j}$ and $\sigma_{jj'}$.

The next approximation we make is to truncate a Hermite basis expansion of the angularly integrated cross section,
\begin{equation}
    \label{eq10:method}
    v_{sa}\sigma_{jj'} = f_{jj'}\sum_{n'',m'',p''}D_{n'',m'',p''}(E_{jj'})\Psi_{n'',m'',p''}(\vec{\xi}) \;\;,
\end{equation}
where we have made use of dependence of $E_k$ on $\vec{\xi}$ through Eqs.~\eqref{eq3:method} and~\eqref{eq1:numver}.
This approximation also implicitly assumes $\vec{v}_{sa}\approx\vec{v}_s$, since the argument of the basis
function is $\vec{\xi}$, hence independent of the ion/atom velocity $\vec{v}_a$.
Linearly factoring $f_{jj'}$ in Eq.~\eqref{eq10:method} would not be possible if we use generalized oscillator
strengths, since $f_{jj'}$ would then depend on the particle momentum transfer, as in Eq.~\eqref{eq8:cros}.
As we will see numerically in Section~\ref{sec:numver}, in the limit $E_{jj'}\ll E_k$ for all level pairs $(j,j')$,
given Eq.~\eqref{eq11:cros} is inversely proportional to $E_{jj'}$ to leading order,
$\log(D_{n'',m'',p''}(E_{jj'}))$ is very close to being linear in $\log(E_{jj'})$.
This permits us to store $D_{n'',m'',p''}(E_{jj'})$ as a set of two-parameter values that fit linear functions
in this log space.
Furthermore, the symmetry in radial velocity of the original kernel implies that $D_{n'',m'',p''}$ is invariant
under permutations of $(n'',m'',p'')$, which permits further savings in data storage.

Incorporating Eq.~\eqref{eq10:method} into the first and second terms of Eq.~\eqref{eq9:method},
\begin{multline}
    \label{eq11:method}
    \frac{\partial C_{n',m',p'}}{\partial t} =
    \sum_{n,m,p}C_{n,m,p}\left[\sum_{n'',m'',p''}f_{jj'}D_{n'',m'',p''}(E_{jj'}) \;\; * \right.\\
    \left\{
    \int f_{a,j}(\vec{v}_a') d^3\vec{v}_a'
    \int\Psi^{n',m',p'}\left(\frac{1}{m_e + M_a}\left(m_e\vec{\xi}'-M_a\vec{u}\oslash\vec{\alpha}\right)\right)
    \Psi_{n'',m'',p''}(\vec{\xi}')\Psi_{n,m,p}(\vec{\xi}') d^3\vec{\xi}'
    \right.\\\left.\left.
    - \int f_{a,j'}(\vec{v}_a) d^3\vec{v}_a
    \int\Psi^{n',m',p'}(\vec{\xi})\Psi_{n'',m'',p''}(\vec{\xi})\Psi_{n,m,p}(\vec{\xi}) d^3\vec{\xi}
    \right\}\right] \;\;,
\end{multline}
where, given the preceding approximation, the integrals over $\beta$-velocity and atom/ion velocity
are separable, hence factored here.
We have also used $\sigma_{j'j}(E_{jj'}) = \sigma_{jj'}(E_{jj'})$, under the assumption that permuting
the initial and final energy levels does not change the cross section structure (statistical weights
from partition functions factoring in the ion distributions, $f_{a,j}$, break the equality for rates,
however). \todo{Verify...}

Given $m_e \ll M_a$, the next approximation we might make is
$\Psi^{n',m',p'}\left(\left(m_e\vec{\xi}'-M_a\vec{u}\oslash\vec{\alpha}\right)/(m_e + M_a)\right)
\approx \Psi^{n',m',p'}(-\vec{u}\oslash\vec{\alpha})$, which gives
\begin{multline}
    \label{eq12:method}
    \frac{\partial C_{n',m',p'}}{\partial t} =
    \sum_{n,m,p}C_{n,m,p}\left[\sum_{n'',m'',p''}f_{jj'}D_{n'',m'',p''}(E_{jj'}) \;\; * \right.\\
    \left\{
    \Psi^{n',m',p'}(-\vec{u}\oslash\vec{\alpha})\int f_{a,j}(\vec{v}_a') d^3\vec{v}_a'
    \int\Psi^{0,0,0}(\xi')\Psi_{n'',m'',p''}(\vec{\xi}')\Psi_{n,m,p}(\vec{\xi}') d^3\vec{\xi}'
    \right.\\\left.\left.
    - \int f_{a,j'}(\vec{v}_a) d^3\vec{v}_a
    \int\Psi^{n',m',p'}(\vec{\xi})\Psi_{n'',m'',p''}(\vec{\xi})\Psi_{n,m,p}(\vec{\xi}) d^3\vec{\xi}
    \right\}\right] \;\;,
\end{multline}
where we have inserted $\Psi^{0,0,0}(\xi')=1$ in the integral, in order to show it is now a
special case of the second pre-scatter integral.
For more generality, we could have expanded
$\Psi^{n',m',p'}\left(\left(m_e\vec{\xi}'-M_a\vec{u}\oslash\vec{\alpha}\right)/(m_e + M_a)\right)$
in terms of the upper-index form of the basis, but this does not add significant formulaic complication
(but it does complicate numerical computation).

The triple-basis integrals in Eq.~\eqref{eq12:method} are separable by dimension in Cartesian velocity
space, resulting in three integrals, where each has an integrand that is a product of three Hermite basis
functions: one upper-index and two lower-index.
For two upper-index and one lower-index Hermite basis, the solution of the integral has been developed
via use of a 3-dimensional generator function by~\cite{askey1999}.
To evaluate the one upper-index, two lower-index integral, we may revisit the generator function
approach given by~\cite{askey1999}, to find a closed-form finite sum that embeds the two upper-index, one
lower-index version of the function.
We provide this derivation in Appendix~\ref{app:B}, and using the result, Eq.~\eqref{eq20:appB},
Eq.~\eqref{eq12:method} becomes
\begin{multline}
    \label{eq13:method}
    \frac{\partial C_{n',m',p'}}{\partial t} =
    \sum_{n,m,p}C_{n,m,p}\left[\sum_{n'',m'',p''}f_{jj'}D_{n'',m'',p''}(E_{jj'}) \;\; * \right.\\
    \left\{
    \Psi^{n',m',p'}(-\vec{u}\oslash\vec{\alpha})\left(\int f_{a,j}(\vec{v}_a') d^3\vec{v}_a'\right)
    T_c(n,0,n'')T_c(m,0,m'')T_c(p,0,p'')
    \right.\\\left.\left.
    - \left(\int f_{a,j'}(\vec{v}_a) d^3\vec{v}_a\right)
    T_c(n,n',n'')T_c(m,m',m'')T_c(p,p,p'')
    \right\}\right] \;\;.
\end{multline}
If we are given the ion distribution $f_{a,j}(\vec{v}_a)$ for all $j$ and fitting coefficients
for $D_{n,m,p}$, we may evaluate the matrix by summing over all pairs $(j,j')$ to get the spectral
formulation of Eq.~\eqref{eq2:awhb}.
Moreover, the ion velocity integrals merely result in the population density for states $j$ and $j'$,
$N_{a,j}$ and $N_{a,j'}$.
The result is
\begin{multline}
    \label{eq14:method}
    \frac{\partial C_{n',m',p'}}{\partial t} =
    \sum_{n,m,p}C_{n,m,p}\Biggl[\sum_{n'',m'',p''}
    \left\{
    \Psi^{n',m',p'}(-\vec{u}\oslash\vec{\alpha})D_{n'',m'',p''}^{(U)}(\vec{x},t)T_c(n,0,n'')T_c(m,0,m'')T_c(p,0,p'')
    \right.\\\left.
    - D_{n'',m'',p''}^{(L)}(\vec{x},t)T_c(n,n',n'')T_c(m,m',m'')T_c(p,p,p'')
    \right\}\Biggr] 
    = \sum_{n,m,p}C_{n,m,p}S_{n,m,p}^{n',m',p'} \;\;.
\end{multline}
where
\begin{subequations}
    \label{eq15:method}
    \begin{gather}
        D_{n,m,p}^{(L)}(\vec{x}, t) 
        = \sum_{(j',j)\in\mathcal{T}}N_{a,j'}(\vec{x}, t)f_{jj'}D_{n,m,p}(E_{jj'}) \;\;, \\
        D_{n,m,p}^{(U)}(\vec{x}, t) 
        = \sum_{(j',j)\in\mathcal{T}}N_{a,j}(\vec{x}, t)f_{jj'}D_{n,m,p}(E_{jj'}) \;\;.
    \end{gather}
\end{subequations}

Similar to $D_{n,m,p}$, the function $T_c(n,n',n'')$ is symmetric under permutation of $(n,n',n'')$.
Furthermore, $D_{n,m,p}$ and $T_c(n,n',n'')$ are spatially invariant functions; the ion distribution
ultimately encodes spatial variation in the spectral scattering matrix.
The spatial invariance and symmetry under index permutation make $D_{n,m,p}$ and $T_c(n,n',n'')$,
as well as the linear fitting representation of $D_{n,m,p}$, very memory efficient
in computational storage, compared to the spectral solution $C_{n,m,p}$.
For instance, considering the basis symmetry, for $(n,m,p)\in\{0,1,2,3,4\}^3$, the number of $C_{n,m,p}$
values to store is 125 per spatial zone, while $T_c(n,m,p)$ results in at most 35 distinct floating point
values, and $D_{n,m,p}$ results in 35 distinct pairs of floating point values (assuming two-parameter
linear fits over atomic transition energy $E_{jj'}$, as discussed).
The $T_c(n,m,p)$ function is also made asymptotically $\sim$50\% sparse for large $(n,m,p)$, by the $(n,m,p)$
parity selection rules discussed in Appendix~\ref{app:B}.

Specifically for Section~\ref{sec:3dkn}, our final main approximation is to assume Maxwellian velocity
dependence of the ion distribution $f_{a,j}(\vec{v}_a)$ and Boltzmann statistics for excitation energy.
This is a LTE assumption about the relative population of the excitation states within an ionization stage.
While we also restrict our calculations to a single ion stage, we note that Saha statistics (hence the
Saha-Boltzmann formulation, for instance as presented by~\cite{mihalas1984}) can be used, without complicating
the above matrix formulae.
For more detail on the integrals using the Maxwellian ion distribution, as well as evaluation of the solid
angle integral of the differential cross section, see Appendix~\ref{app:A}.

\subsection{A general derivation outline}

To end this section, we note that the calculation procedure described above may translate to a more
general derivation sequence, applicable to different types of collision kernels.
These general steps are outlined here.
\begin{enumerate}
    \item Identify the optimal coordinate system of velocity space to spectrally expand
    functional forms of the solution distribution and the differential cross section.
    \item Choose one or more spectral basis in the velocity coordinate system, based on physics
    (for example Gaussian weighting function for inner products, for natural scale-bridging, or relativity-compatible
    functions such as the Maxwell-J\"{u}ttner distribution) and physical regime,
    and expand both the distribution function and cross section in terms of the basis functions.
    \item Determine if parametric fits (for example, in this work over atomic transition energy)
    are applicable to the fundamental cross section coefficients ($D_{n,m,p}$).
    \begin{enumerate}
        \item Store the spatially invariant form of the parameterized coefficients if possible
        (for instance, excluding integration of the coefficient over the atom/ion distribution).
        \item Determine any index symmetries to further compress the $D_{n,m,p}$ array.
    \end{enumerate}
    \item Incorporate the expansions into the collision kernel relating the distribution to the
    differential cross section.
    \item Use the principle of micro-reversibility to (possibly) simplify evaluation of the post-scatter term.
    \item Determine if approximations, or an expansion, is possible to separate kernel integrals
    by species (for example, the separation of the integral over ion velocity above).
    \item Integrate and use basis orthogonality to arrive at a spectral matrix equation.
    \item If possible, use closed-form expressions for basis function triple-products~\citep{askey1999}.
        \begin{enumerate}
            \item If the cross section and distribution use the same basis functions, it may be possible
            to follow the generator function procedure given by~\cite{askey1999} to find a closed form.
            \item If the cross section uses a different basis expansion (for example, Legendre polynomials),
            it may be possible to use a Gram-Schmidt orthonormalization procedure to express the polynomial
            factor of one basis in terms of the polynomial (upper-index) factor of the other.
            \item Determine any index symmetries to further compress pre-computation of the triple
            product function.
            \item Identify index selection rules that may increase sparsity of triple product function evaluation.
        \end{enumerate}
\end{enumerate}

\section{Numerical Verification of Spectral Collision Matrix} \label{sec:numver}

In this section, we numerically verify different aspects of the computation of the spectral collision matrix
$S_{n,m,p}^{n',n',p'}$, in order to motivate the proof-of-principle 3D KN simulation presented in Section
\ref{sec:3dkn}.
First, we numerically verify the correctness of the compact triple Hermite products involving one upper-index
basis and two lower-index basis functions.
Next, we examine spectral expansions of the uncollided $\beta$-particle distribution (or the emitted
$\beta$-particle distribution) and the fast-particle collision kernel.
We then verify the two-parameter coefficient fits used for the closed-form expansion of the cross section match
the direct calculation over different values of atomic transition energy.
Finally, we compare direct numerical integration of Eq.~\eqref{eq9:method} with the evaluation of
Eq.~\eqref{eq13:method}.
For all evaluations of the kernel or cross section, we assume a collision parameter of $\epsilon=10^{-3}$ in
Eq.~\eqref{eq20:awhb}, which corresponds to a minimum collision angle of $\sim2.5^{\circ}$.
Importantly, in this and the following section, for expanding the $\beta$-particle emission and cross sections,
we initially transform the kinetic energy into radial velocity using the relativistic formula,
\begin{equation}
    \label{eq1:numver}
    |\vec{v}| = c\sqrt{1 - \left(\frac{1}{E_k/(m_ec^2)+1}\right)^2} \;\;,
\end{equation}
in order to systematically restrict the velocity domain to a more causal region of velocity space.
However, the Hermite basis still extends beyond $|\vec{v}|=c$.

In Table~\ref{tab1:numver}, we compare the closed form of the compact triple Hermite products
(Appendix~\ref{app:B}, Eq.~\eqref{eq20:appB}) to a direct 1024-point Midpoint Rule numerical
integration over the non-dimensional velocity domain $\xi\in(-4,4)$.
We see very good agreement for all values, but an increased discrepancy at higher order basis
functions.
This seems to be due to the integral bounds in $\xi$ not being extended far enough for accuracy
in the direct numerical integration (the bounds are supposed to be indefinite, $\xi\in(-\infty,\infty)$);
as the bounds are increased the Midpoint Rule implementation
comes into closer agreement with the closed form across modes.
\begin{table}[]
    \centering
    \begin{tabular}{|l|r|r|r|r|}
        \hline
         & $T_c(0,0,0)$ & $T_c(1,1,1)$ & $T_c(1,2,3)$ & $T_c(3,1,2)$ \\
         & $T_c(2,2,2)$ & $T_c(4,0,0)$ & $T_c(0,4,0)$ & $T_c(4,4,4)$ \\
        \hline\hline
        Midpoint Rule 
        & 0.3989422804014322 & 0.0 & 0.043186768679 & 0.043186768679 \\
        & 0.017630924480 & 0.0610753139878 & 0.0610753139878 & 0.001431449 \\
        \hline
        Closed Form 
        & 0.3989422804014327 & 0.0 & 0.043186768684 & 0.043186768684 \\
        & 0.017630924486 & 0.0610753139879 & 0.0610753139879 & 0.001431452 \\
        \hline
    \end{tabular}
    \caption{
        Numerical values of compact triple Hermite products using 1024-point Midpoint Rule
        over $\xi\in(-4,4)$ (first row) and the closed form presented in Appendix~\ref{app:B}.
        The function is symmetric under permutation of the three index arguments, corresponding to
        each basis order involved in the inner product.
    }
    \label{tab1:numver}
\end{table}

\begin{figure}
    \centering
    \includegraphics[height=64mm]{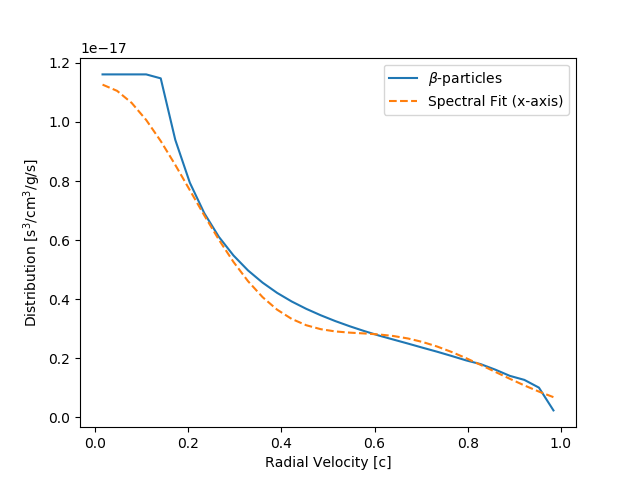}
    \includegraphics[height=64mm]{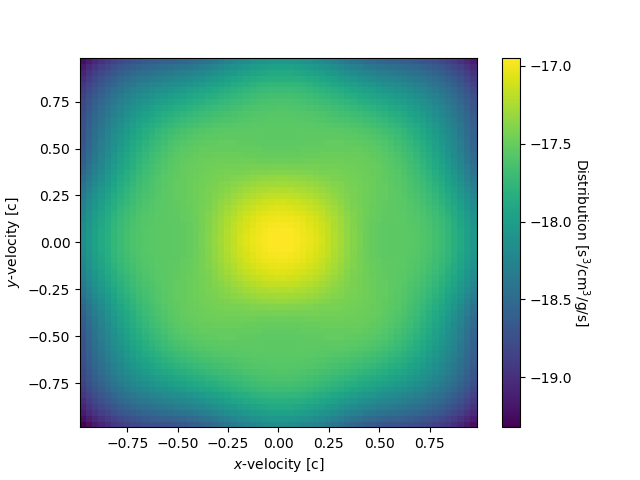}
    \caption{
        Left: Reference uncollided $\beta$-particle distribution (blue) and line-out
        of corresponding spectral reconstruction over 9x9x9 velocity basis functions (dashed orange),
        versus radial velocity.
        The original distribution is calculated from 0.2$c$, and is set to a constant
        for lower velocity.
        Right: Base-10 log of reconstructed uncollided $\beta$-particle distribution versus
        $xy$-plane of velocity space.
    }
    \label{fig1:numver}
\end{figure}

Turning to the spectral reconstruction of the uncollided $\beta$-particle distribution and scatter
angle-integrated collision kernel, some numerical experimentation indicates $\alpha_x=\alpha_y=\alpha_z=0.5c$
and $u_x=u_y=u_z=0$ are reasonable basis parameters for both.
Figure~\ref{fig1:numver} has a plot of the uncollided $\beta$-particle distribution (solid blue),
calculated from $\beta$-particle emission spectrum \todo{cite paper}, and a line-out of the
spectrally reconstructed profile (dashed orange), using Eq.~\eqref{eq2:method} with $n$, $m$, $p$
each ranging from 0 to 8 in Hermite basis order.
We see some oscillation in the fit, which resembles the well-known Gibbs phenomenon (the ringing
artefacts are near high curvature in the profile).
The fit is poor for radial velocity below $\sim0.2c$, where the original distribution is set
to a constant.
This also is a range where the validity of the fast-particle quantum approximation may start
to break down~\citep{inokuti1971}.

\begin{figure}
    \centering
    \includegraphics[height=64mm]{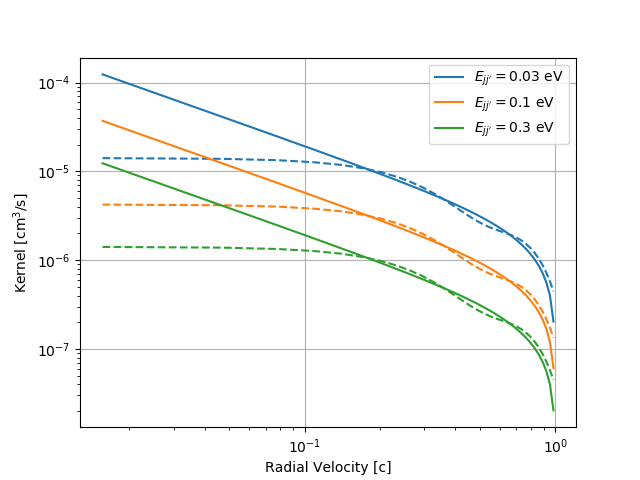}
    \includegraphics[height=64mm]{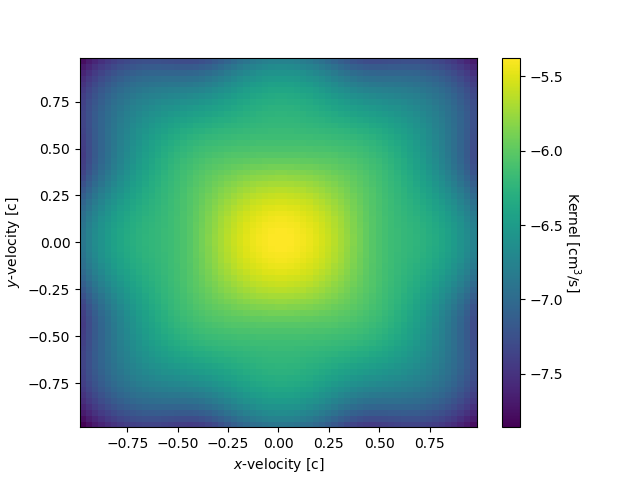}
    \caption{
        Left: Reference one-line scatter angle-integrated kernel (solid) and line-out of
        corresponding spectral reconstruction kernel over 9x9x9 velocity basis functions (dashed),
        for atomic transition energies of 0.03, 0.1, and 0.3 eV.
        Right: Base-10 log of reconstructed one-line scatter angle-integrated kernel versus
        $xy$-plane of velocity space.
    }
    \label{fig2:numver}
\end{figure}

The left panel of Fig.~\ref{fig2:numver} shows the original (solid) and reconstructed (dashed)
kernel, where the reconstructed kernel is over the same 9x9x9 basis as in Fig.~\ref{fig1:numver}
(again with $\alpha_x=\alpha_y=\alpha_z=0.5c$ and $u_x=u_y=u_z=0$).
Some oscillation can be seen in the reconstructed kernel as well, and coincidentally the fit
is also poor at radial velocity below $\sim0.2c$, where error from the original kernel approximation
may start to be significant.
Furthermore, in the right panel of Fig.~\ref{fig2:numver} is a plot of the base-10 log of the
reconstructed kernel over the $xy$-plane in velocity space.
While the original kernel is radially symmetric, the reconstructed kernel shows artifacts from
fitting over a Cartesian basis of Hermite functions.

\begin{figure}
    \centering
    \includegraphics[height=64mm]{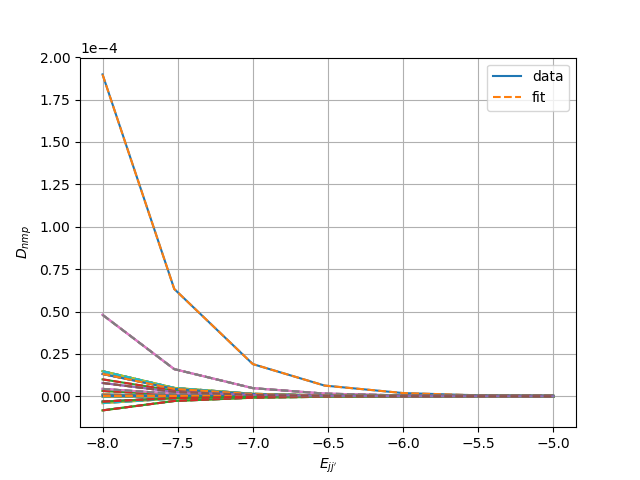}
    \includegraphics[height=64mm]{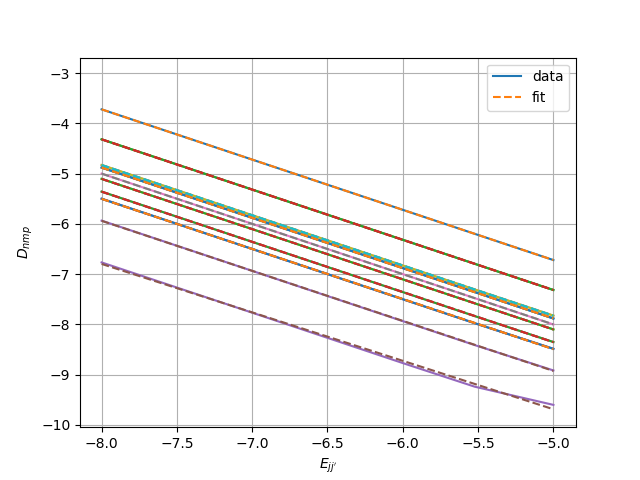}
    \caption{
        $D_{n,m,p}$ versus parameterized atomic energy transition $E_{jj'}$, for a few selected 
        values of $(n,m,p)$, comparing direct evaluation of the expansion coefficients (solid)
        to linear fits in log-log space (dashed).
        Left: $D_{n,m,p}$ versus log base-10 of $E_{jj'}$ in MeV (0.001 to 10 eV).
        Right: log base-10 $D_{n,m,p}$ versus log base-10 of $E_{jj'}$.
    }
    \label{fig3:numver}
\end{figure}

Figure~\ref{fig3:numver} has $D_{n,m,p}$ versus a parameterized transition energy $E_{jj'}$, for a few
selected values of $(n,m,p)$, comparing direct evaluation of the expansion coefficients to linear fits 
in log-log space.
We observe that the linear fits in log-log space do well to capture the dependence of each $D_{n,m,p}$
term for the range of atomic transition energies considered, $\sim$0.001 to 10 eV.
For subsequent calculations involving construction of the spectral matrix, we use these fits to $D_{n,m,p}$
(which are particularly important for Section~\ref{sec:3dkn}).

The Hermite basis reconstructions in Figs.~\ref{fig1:numver} and~\ref{fig2:numver} suggest
orders 0 to 8 in each dimension may furnish reasonable accuracy, notwithstanding the approximations
already made in the kernel and $\beta$-particle emission.
We now verify that these Hermite basis orders are sufficient for accuracy when incorporated into
the pre-scatter portion of Eq.~\eqref{eq12:method}.
To do so, we compare direct numerical integration of Eq.~\eqref{eq9:method} to the evaluation of
Eq.~\eqref{eq13:method}, for a single line with oscillator strength $f_{jj'}=1$ and transition
energy $E_{jj'}=0.1$ eV.
Table~\ref{tab2:numver} has numerical values for some entries of the spectral scattering matrix, for
direct integration by the Midpoint Rule on a $64^3$ point velocity domain and using Eq.~\eqref{eq13:method},
with either 5x5x5 or 9x9x9 Hermite basis functions.
Also in Table~\ref{tab2:numver} is relative error, as a fraction of the direct numerical integral.
The matrix elements which are very close numerically to 0 have high error, but these terms will not contribute to
the solution.
Otherwise, for the significant entries, we see the low order values are in close agreement but along
rows/columns of the matrix the error increases with higher disparity between modes, at $\sim$28\%
for the $S_{0,0,0}^{4,0,0}$ term using 5x5x5 basis functions.
However, in using Eq.~\eqref{eq13:method}, we do not have to restrict the innermost sum to 5x5x5, even when
simulating $C_{n,m,p}$ modes only up to 5x5x5.
If we permit this innermost sum to go instead to 9 in each velocity dimension, we obtain results
with systematically lower error relative to direct integration ($\lesssim10$\% for non-trivial modes).
These results indicate that a 9x9x9 basis truncation of the innermost sum can accurately integrate the matrix terms, consistent with the accuracy of the profile of the truncated kernel expansion shown
in Fig.~\ref{fig2:numver}.
\begin{table}[]
    \centering
    \begin{tabular}{|l|r|r|r|r|r|}
        \hline
         & $S_{0,0,0}^{0,0,0}$ & $S_{0,0,0}^{1,0,0}$ & $S_{0,0,0}^{2,0,0}$ & $S_{0,0,0}^{3,0,0}$ & $S_{0,0,0}^{4,0,0}$ \\
         & $S_{2,0,0}^{0,0,0}$ & $S_{2,0,0}^{1,0,0}$ & $S_{2,0,0}^{2,0,0}$ & $S_{2,0,0}^{3,0,0}$ & $S_{2,0,0}^{4,0,0}$ \\
         & $S_{4,0,0}^{0,0,0}$ & $S_{4,0,0}^{1,0,0}$ & $S_{4,0,0}^{2,0,0}$ & $S_{4,0,0}^{3,0,0}$ & $S_{4,0,0}^{4,0,0}$ \\
        \hline\hline
        Midpoint Rule, Eq.~\eqref{eq9:method}
        & -1.064612e-06 & -2.244660e-24 & 3.773394e-07 & -3.568146e-24 & -1.838709e-07 \\
        & 3.773394e-07 & 1.754513e-24 & -4.477249e-07 & 1.501545e-25 & 3.166727e-07 \\
        & -1.838709e-07 & 2.815265e-25 & 3.166727e-07 & 5.027491e-25 & -3.093139e-07 \\
        \hline
        Eq.~\eqref{eq13:method} (5x5x5 basis)
        & -1.035940e-06 & 8.427530e-24 & 3.404054e-07 & -4.664180e-24 & -1.332969e-07 \\
        & 3.404054e-07 & 3.839731e-24 & -3.996374e-07 & 2.175768e-24 & 2.853456e-07 \\
        & -1.332969e-07 & -4.118894e-24 & 2.853456e-07 & 1.544883e-24 & -2.920202e-07 \\
        \hline
        Eq.~\eqref{eq13:method} (9x9x9 basis)
        & -1.059921e-06 & 5.339311e-24 & 3.688633e-07 & -6.485463e-26 & -1.696097e-07 \\
        & 3.688633e-07 & 7.438594e-24 & -4.320753e-07 & 1.294707e-24 & 2.999444e-07 \\
        & -1.696097e-07 & -5.558943e-24 & 2.999444e-07 & 1.526768e-24 & -2.977645e-07 \\
        \hline\hline
        Error (5x5x5 basis)
        & 0.02693188 & 4.75447952 & 0.09788005 & 0.30717185 & 0.27505168 \\
        & 0.09788005 & 1.1884882 & 0.10740412 & 13.4901951 & 0.0989258 \\
        & 0.27505168 & 15.63057297 & 0.0989258 & 2.07287074 & 0.05590987 \\
        \hline
        Error (9x9x9 basis)
        & 0.0044063 & 3.37867249 & 0.0224628 & 0.981824 & 0.07756094 \\
        & 0.0224628 & 3.23969158 & 0.03495361 & 7.62249883 & 0.0528252 \\
        & 0.07756094 & 20.74571843 & 0.0528252 & 2.03683885 & 0.03733877 \\
        \hline
    \end{tabular}
    \caption{
        Numerical values of the pre-scatter portion of the spectral collision matrix, and relative errors
        as a fraction of numerical integral.
        The top three rows are from direct numerical integration of Eq.~\eqref{eq9:method} over a
        64$^3$ point velocity domain using the Midpoint Rule.
        The next three rows are from evaluating Eq.~\eqref{eq13:method} using compact triple
        Hermite products and the fitted data coefficients $D_{n,m,p}$, using 5x5x5 basis functions.
        The next three rows are again for Eq.~\eqref{eq13:method}, but using 9x9x9 basis functions.
        The bottom six rows are the relative error between the Midpoint Rule and
        Eq.~\eqref{eq13:method}, as a fraction of the Midpoint Rule.
    }
    \label{tab2:numver}
\end{table}

\begin{figure}
    \centering
    \includegraphics[height=64mm]{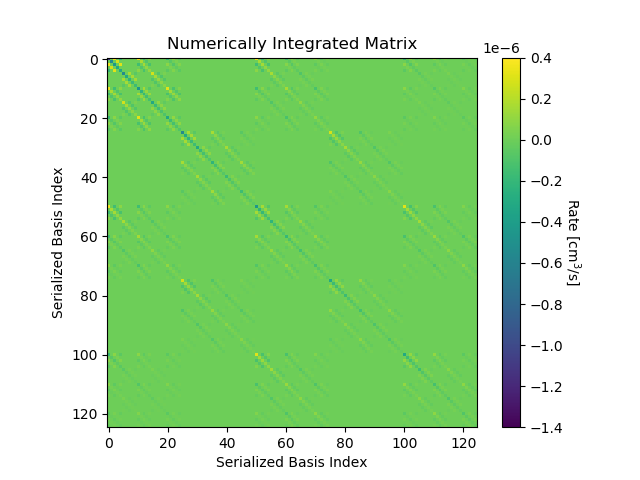}
    \includegraphics[height=64mm]{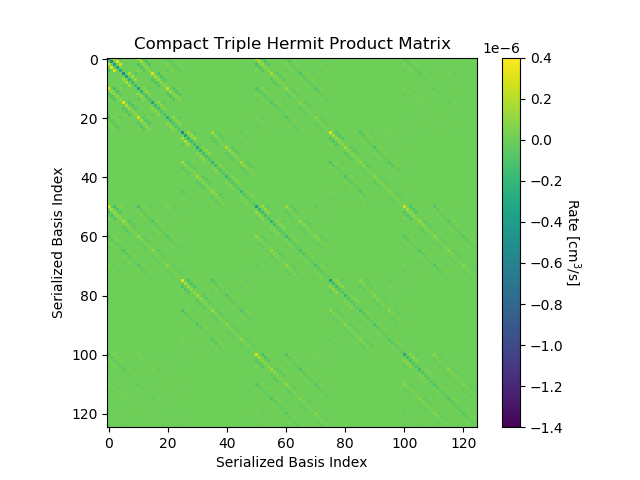}
    \caption{
        Spectral collision matrix elements versus matrix indices serialized over Hermite
        basis order.
        Left: a direct numerical integration over velocity space of Eq.~\eqref{eq9:method}
        using the Midpoint Rule on a $64^3$ velocity grid.
        Right: evaluation of Eq.~\eqref{eq13:method} for each matrix entry.
    }
    \label{fig5:numver}
\end{figure}

In what follows, we restrict our attention to the pre-scatter matrix, since the post-scatter
matrix does not complicate the analysis.
Again we assume a single line, a pre-scatter ion population density of 1 (effectively), an
atomic transition energy of 0.1 eV, and an oscillator strength of 1.
In Fig.~\ref{fig5:numver}, we show the full matrix for a 5x5x5 velocity basis functions versus serialized
matrix indices.
The left panel has the direct numerical integral of Eq.~\eqref{eq9:method} for each matrix entry,
again using a Midpoint rule over 64$^3$ points in velocity space, and the right panel
uses Eq.~\eqref{eq13:method} for each matrix entry (hence using the closed form compact triple
Hermite product functions).
Even over the coarse velocity space grid, the cost of directly numerically integrating the
pre-scatter matrix is significantly higher than using the compact triple Hermite product functions:
 1743 seconds for the numerical integration and 2.2 (5.1)
seconds for the 5x5x5 (9x9x9) innermost sum (using Python/NumPy on a single CPU).
This time comparison is in fact for a sub-optimal implementation of the compact triple Hermite
functions, where they are reevaluated for each instance they are invoked, rather than simply
pre-computed.
Moreover, the structure of the matrix matches between the two methods (for this kernel, we observe
that much of the qualitative structure comes from the triple-products of the Hermite basis function,
which can be seen by setting all the $D_{n,m,p}$ values to a constant and comparing to the matrix using
detailed atomic data).

\todo{Other numerical tests?}


\section{Simplified 3D KN ejecta thermalization trial calculation} \label{sec:3dkn}

In Section~\ref{sec:numver} we verify the important steps of computing the terms of the
spectral matrix $S_{n,m,p}^{n',m',p'}$, which linearly couple together modes of the solution
$C_{n,m,p}$.
We now apply the 9x9x9 basis functions used in Section~\ref{sec:numver}, with the parameters
$\vec{\alpha}=0.5$ and $\vec{u}=0$, to a proof-of-principle KN model in 3D Cartesian spatial geometry.
We use the MASS-APP code base, which is a spectral Hermite solver for the Vlasov-Maxwell-Boltzmann equations.
The MASS-APP code non-dimensionalizes the equations following Eq.~\eqref{eq0:appC}; we leverage the
reference plasma electron oscillation frequency in Eq.~\eqref{eq0:appC} to match the expansion time scale
of the KN, setting it to $5\times10^{-5}$ radians/s.
The physical length scale of KNe from 1 day to a week is $\sim 10^{14}-10^{15}$ cm, which implies the
reference plasma oscillation frequency furnishes a non-dimensional length scale of O(1).
Consequently, we set the 3D spatial domain to a $[-1,1]\times[-1,1]\times[-1,1]$ non-dimensional cube.

The KN model has an axisymmetric ejecta with a toroidal (T) component superimposed with lobed
(P; ``peanut'') component.
The formula for the ion density is derived from the Cassini oval approach of~\cite{korobkin2021},
\begin{equation}
    \label{eq1:3dkn}
    N_a(\bar{x},\bar{y},\bar{z}) = N_{a,0}
    \begin{cases}
        (1 - \bar{r}^4 + 4\bar{z}^2 - 2\bar{r}^2)^3
        \;\;,\;\;\text{(T)}\;\;,\\
        (1.5 - \bar{r}^4 - 4\bar{z}^2 + 2\bar{r}^2)^3
        \;\;,\;\;\text{(P)}\;\;,
    \end{cases}
\end{equation}
where $N_{a,0}$ is number density, $\bar{x}$, $\bar{y}$, and $\bar{z}$ are scaled non-dimensional
spatial Cartesian coordinates, and $\bar{r}^2=\bar{x}^2+\bar{y}^2+\bar{z}^2$.
We set the scaling to 2, so $\bar{\vec{r}} = 2\tilde{\vec{r}}$, where the components of $\tilde{\vec{r}}$
range from -1 to 1 and are non-dimensional in the form of Eq.~\ref{eq0:appC}.
Considering the typical homologous approximation for KN (or supernova) ejecta, we see that the
non-dimensionalization procedure implies,
\begin{equation}
    \tilde{\vec{x}} = \frac{\vec{v}_{\rm exp}}{c}\tilde{t}_{\rm exp} \;\;,
\end{equation}
where $\vec{v}_{\rm exp}$ is the bulk expansion velocity of the ejecta, and $\tilde{t}_{\rm exp}$ is the
non-dimensional time elapsed since the merger event.
A non-dimensional expansion time of 5 then corresponds to $10^5$ seconds of physical time, or about 1 day,
which translates to a physical expansion velocity of 0.2$c$ for $\tilde{x}=1$, $\tilde{z}=\tilde{y}=0$.
This configuration effectively sets the ejecta velocity scale between the two components to be comparable
\citep{korobkin2021}.

\begin{figure}
    \centering
    \includegraphics[height=64mm]{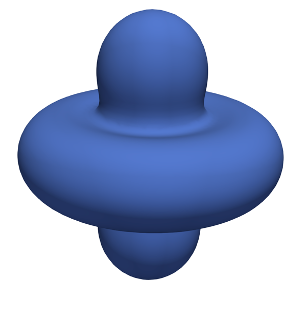} \\
    \includegraphics[height=64mm]{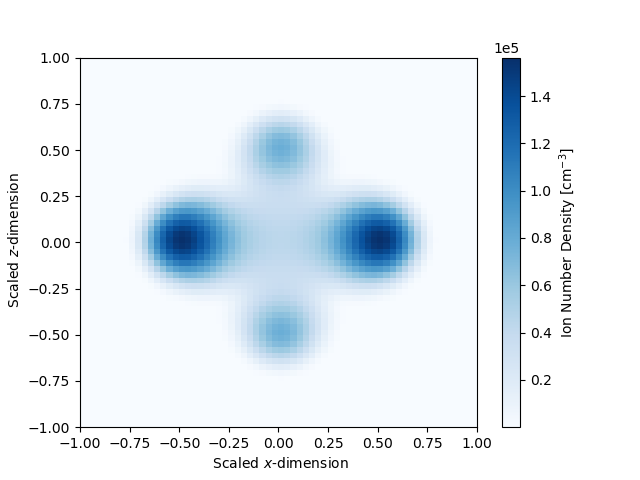}
    \includegraphics[height=64mm]{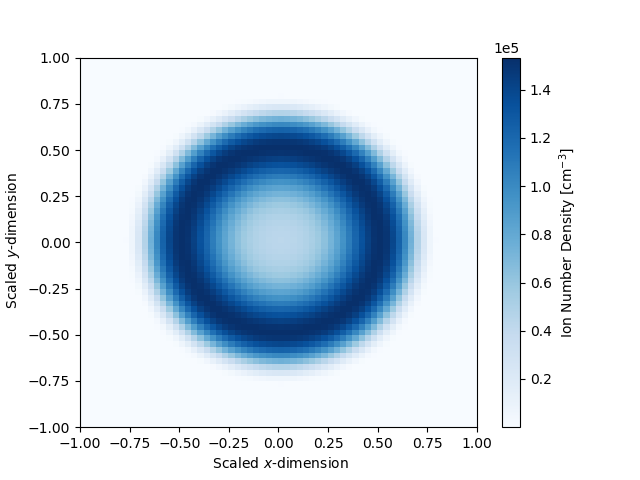}
    \caption{
        \textbftog{Top:} Isocontour of 2D axisymmetric KN ejecta morphology from~\cite{korobkin2021}
        (TP morphology) embedded in 3D Cartesian space, showing torus and axial wind components\textbftog{,
        using Eq.~\eqref{eq1a:3dkn} with parameters described in Section~\ref{sec:3dkn}}.
        \textbftog{Bottom: ion number density of the same profile in $zx$ (left) and $xy$ (right) planes.}
    }
    \label{fig1:3dkn}
\end{figure}

We set the reference number density $N_{a,0}$ for the profile to $10^4$ cm$^{-3}$, which is very low, corresponding
to an ejecta mass of $5\times10^{-5}$ solar masses.
\textbftog{This choice approximates the fraction of mass of Neodymium (Nd) in a more realistic total (see, for instance,
Table 1 of~\cite{even2020}); Nd has a significant impact on the photon opacity~\citep{even2020}.}
The total density profile is a sum over the components where each component imposes a minimum background
density of $N_{a,\min}=10^{-2}$ cm$^{-3}$,
\begin{equation}
    \label{eq1a:3dkn}
    N_{a,{\rm tot}}(\bar{x},\bar{y},\bar{z}) = 
        \max(N_{a,\min}, N_{a,0}(1 - \bar{r}^4 + 4\bar{z}^2 - 2\bar{r}^2)^3)
        + \max(N_{a,\min}, N_{a,0}(1.5 - \bar{r}^4 - 4\bar{z}^2 + 2\bar{r}^2)^3) \;\;.
\end{equation}
Figure~\ref{fig1:3dkn} has isocontour \textbftog{(top panel) and $zx$, $xy$-plane (bottom left and right panels)}
plots of the ion density given by Eq.~\eqref{eq1:3dkn}, showing the shape of the combined ejecta components.

The ion temperature is assumed to be isothermal, or uniform in space (consistent with thermal electron
temperature at late time in some LTE two-temperature KN simulations), and we set it to 0.1 eV.
We make an extreme assumption that the entire ejecta is singly ionized Nd, and use
energy levels ($E_{j}$), statistical weights ($g_j$), and oscillator strengths $f_{jj'}$ from
the LANL suite of atomic physics codes~\citep{fontes2015b}.
We neglect oscillator strengths below $10^{-3}$, leaving a total of 6888 levels, connected by
375026 lines.
Within the singly ionized Nd stage, we determine the excitation levels with the Boltzmann factor
and partition function,
\begin{equation}
    \label{eq2:3dkn}
    \int f_{a,j}(\vec{v})d^3\vec{v}_a = N_{a,j} 
    = \left(\frac{g_je^{-E_{j0}/T_a}}{\sum_{j'}g_{j'}e^{-E_{j'0}/T_a}}\right)N_a \;\;,
\end{equation}
where $T_a$ is the ion temperature.
Equation~\eqref{eq2:3dkn} is an assumption of LTE in the excitation states of the ion.

We simulate the model with 10 uniform time steps and 64$^3$ spatial points over one second of physical time,
using an initial condition for the $\beta$-particle spectrum from Fig.~\ref{fig1:numver},
proportionally scaling with ejecta density $N_a$ to account for higher $\beta$-emission rates at higher
ion densities.
The initial electromagnetic field is set to 0 everywhere ($\vec{E}=\vec{B}=0$).
On the AMD Rome EPYC 7H12 CPU partition of the HPC system Chicoma at LANL, the simulation took 1.4 hours
on 256 cores (we recalculate the matrix entries each time step despite keeping the ion temperature and
density as constant in time).

\begin{figure}
    \centering
    \includegraphics[height=64mm]{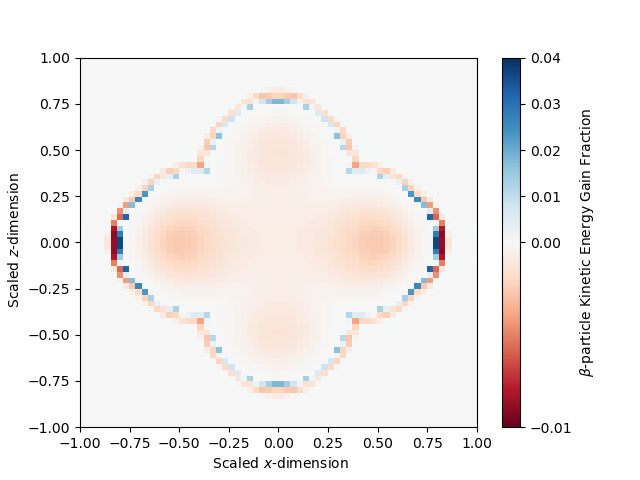}
    \includegraphics[height=64mm]{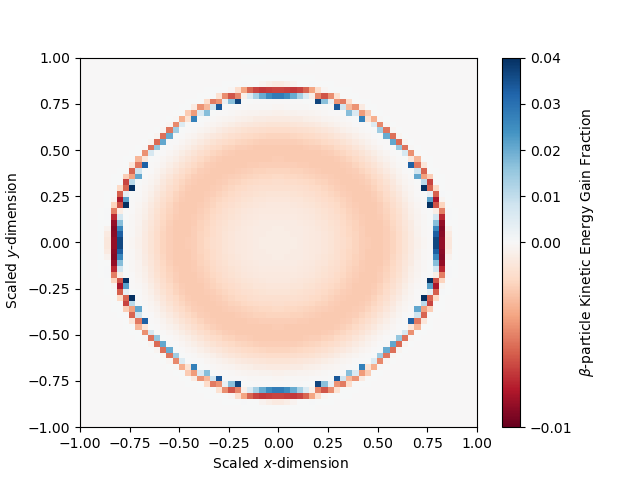}
    \caption{
        Fraction of kinetic energy gain (blue) and loss (red) from $\beta$-particle field 
        in $zx$ (left) and $xy$ (right) planes, for the proof-of-principle KN problem described
        in Section~\ref{sec:3dkn}.
    }
    \label{fig2:3dkn}
\end{figure}

Figure~\ref{fig2:3dkn} has the kinetic energy gain fraction at 1 second, relative to the initial conditions,
in the $zx$ (left) and $xy$ (right) spatial planes.
From these plots, we see the kinetic energy loss is enhanced in regions of high density in the ion field
as expected, but we also observe a thin layer at the edge of the ejecta where a large fractional gain
and loss occur.
This effect may be attributable to a transitional region where the ion density is low enough that flux between
zones begins to dominate over the collision matrix. \todo{Verify...}
We observe in this one second time scale that $\sim$0.3\% ($\sim$0.13\%) of the initial kinetic energy in the
$\beta$-particle field is lost in density regions near the peak ion density of the torus (peanut) component.

\begin{figure}
    \centering
    \includegraphics[height=64mm]{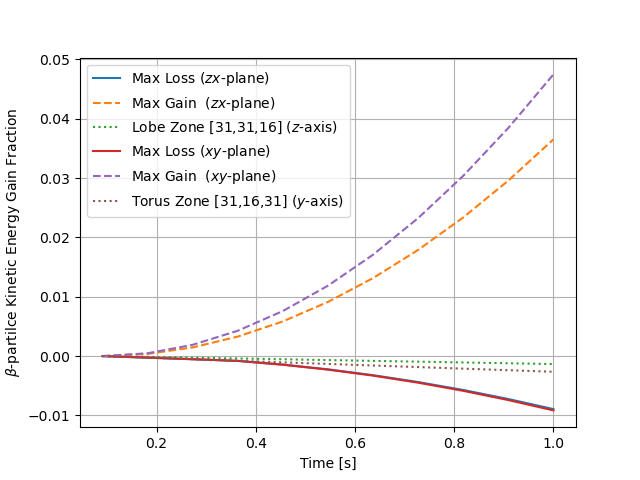}
    \includegraphics[height=64mm]{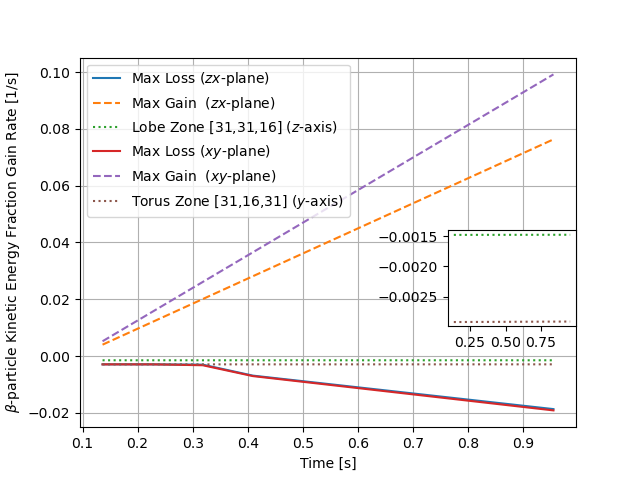}
    \caption{
        Left: Fractional kinetic energy max gain (dashed), max loss (solid), and a selected
        torus/peanut spatial coordinate versus time for the $zx$ and $xy$-planes.
        Right: Corresponding rate of change in fractional kinetic energy, with inset for
        torus and lobe rates.
    }
    \label{fig3:3dkn}
\end{figure}

The left panel of Fig.~\ref{fig3:3dkn} has fractional kinetic energy max gain (solid), max loss (dashed),
and a selected torus/peanut zone versus time for the $zx$ and $xy$-planes.
We see that the fractional kinetic energy losses and gains near the outer edges are higher than losses in the
torus and peanut lobes by a factor of a few and order of magnitude, respectively.
It should be noted that these are spatially local fractional values; the energy lost in the torus and
peanut lobe is orders of magnitude higher (the fractional metric happens to capture local behavior better,
for instance exposing the effect at the ejecta boundary layer).
The right panel of Fig.~\ref{fig3:3dkn} has the corresponding rate of change (time derivative) of the
left panel data.
The rate of change in the fraction of kinetic energy lost to the ions, relative to the original amount
in the spatial zone, is nearly constant over the second for torus and peanut lobe zones, at approximately
0.3\% s$^{-1}$ and \textbftog{0.15\% s$^{-1}$} respectively, but grows in magnitude for the ejecta boundary layers.
Assuming
\begin{equation}
    \mathit{f}_{\beta,{\rm therm}}(t)\dot{\epsilon}_k(t) \approx
    \dot{\mathit{f}}_{\beta,{\rm therm}}(\epsilon_k(t)-\epsilon_k(0)) \;\;,
\end{equation}
where $\mathit{f}_{\beta,{\rm therm}}$ is the thermalization fraction and $\epsilon_k$ is the kinetic
energy emitted by time $t$, and given that we assumed an emissivity for the initial condition,
such that $\epsilon_k(t)-\epsilon_k(0) = \dot{\epsilon}_k(t)$, the rate of change in the fractions
result in $\mathit{f}_{\beta,{\rm therm}}=0.003$ (T) or \textbftog{0.0015} (P).

We may estimate the magnitude of the effect of excitation collision scattering angles $\geq2.5^{\circ}$ relative
to total $\beta$-thermalization using the semi-analytic formulae of~\cite{barnes2016}, in particular their
Eqs.~20 and 32, reproduced here for convenience,
\begin{subequations}
    \label{eq4:3dkn}
    \begin{gather}
        t_{\beta,{\rm ineff}} = 7.4\left(\frac{E_k}{0.5\text{ MeV}}\right)^{-1/2}
        \left(\frac{M_{\rm ej}}{5\times10^{-3}\text{ M}_{\odot}}\right)^{1/2}
        \left(\frac{\vec{v}_{\rm exp}}{0.2c}\right)^{-3/2} \text{days} \;\;, \\
        \mathit{f}_{\beta,{\rm therm}}(t) = \frac{\ln(1+2(t/t_{\beta,{\rm ineff}})^2)}{2(t/t_{\beta,{\rm ineff}})^2}
        \;\;,
    \end{gather}
\end{subequations}
where $t_{\beta,{\rm ineff}}$ is the time scale to inefficient thermalization, which happens to be about
where the peak of the KN transient matches the thermalization time scale~\citep{barnes2016}, and
$\mathit{f}_{\beta,{\rm therm}}$ is again the thermalization fraction at time $t$, which multiplies the
bare emission rate.
Using the dimensional values of our 3D model, with \textbftog{$t=10^5$ s and} $\vec{v}_{\rm exp} = 0.1c$,
and obtaining an average
uncollided $\beta$-particle energy of 0.3 MeV from integrating $\int E_k f_s(E_k)dE_k / \int f_s(E_k)dE_k$,
we see the inefficiency time scale and thermalization fraction for the total $\beta$-particle interaction
should be roughly \textbftog{2.7} days and \textbftog{0.85}, respectively.
Compared to \textbftog{0.85}, the estimates of 0.003 (T) and \textbftog{0.0015} (P) are much lower, corresponding to
$\sim$0.03 days and $\sim$0.02 days inefficiency time scales, respectively (using the Newton-Raphson method
on Eq.~\eqref{eq4:3dkn}b).
This indicates excitation scattering at large angle is a sub-dominant but not \textbftog{vanishingly small}
mechanism for energy transfer to the ions.
Incorporating large-angle ionization \textbftog{and free electron Coulomb collisions} may increase the large-angle
contribution to thermalization as well \textbftog{\citep{barnes2021}}.
Alluded to earlier, a significant caveat to this numerical evaluation is that the fast-particle approximation
has been built into our derivation for large angle scatters, which for lower energy $\beta$-particles may
become invalid.

\begin{figure}
    \centering
    \includegraphics[height=64mm]{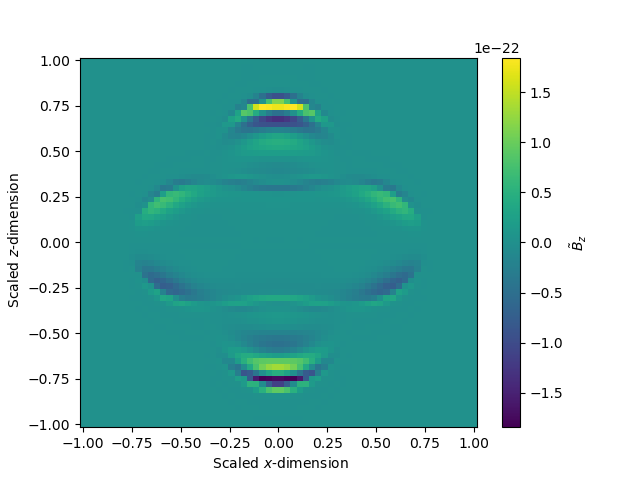}
    \includegraphics[height=64mm]{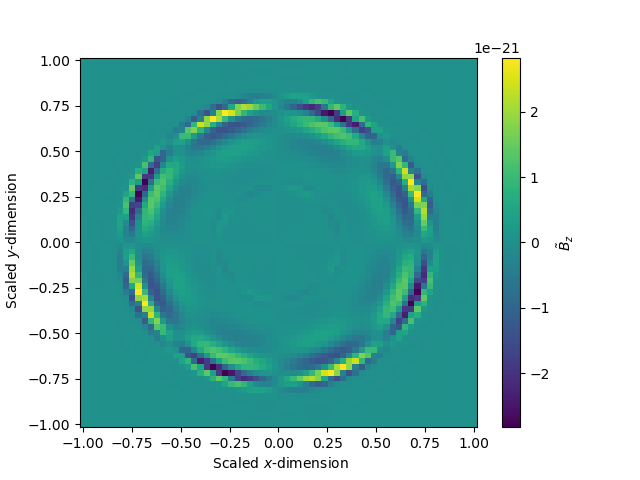}
    \caption{
        Non-dimensional $z$-component of magnetic field at 1 second
        in $zx$ (left) and $xy$ (right) planes, for the proof-of-principle KN problem 
        described in Section~\ref{sec:3dkn}.
    }
    \label{fig4:3dkn}
\end{figure}

Given that the Maxwell equations are also being solved, and we started the simulation with
$\vec{E}=\vec{B}=0$, it may be of interest to see the structure of the magnetic field after
one second.
Figure~\ref{fig4:3dkn} has the $z$-component of the non-dimensional magnetic field
in the $zx$ (left) and $xy$ (right) spatial planes.
The non-dimensional magnetic field is very low, and nearly 0 everywhere except the edges, again
indicating a region where particle flux, and hence current, is important.
The non-zero portions show antisymmetry under reflection through the $xy$-plane, consistent with the
0-divergence condition of the magnetic field.
The alternating field in the $xy$-plane suggests the formation of a very weak long-wavelength
electromagnetic wave near the surface of the toroidal ejecta.
We can see that these values of $\tilde{B}_z$ are subdominant to the collision matrix by examining $C_{0,1,0}$,
which corresponds to particle momentum in the $y$-direction: the time rate of change of $C_{0,1,0}$ is
proportional to $\tilde{B}_zC_{1,0,0}$ through the Lorentz force (see Eq.~\eqref{eq1:appC}), but
it is also proportional to $\sum_{n',m',p'}S_{0,1,0}^{n',m',p'}C_{n',m',p'}$, where the dominant
values of $S_{0,1,0}^{n',m',p'}$ are $\sim$14 orders of magnitude larger than $\tilde{B}_z$.
As a C-coefficient, the electric field is similarly orders of magnitude lower than the dominant
collision matrix elements, but only by $\sim$4 orders of magnitude.

\section{Conclusions} \label{sec:conclude}

We have formulated and implemented a preliminary spectral evaluation of the
fast-particle atomic excitation kernel presented by~\cite{inokuti1971}.
The resulting spectral collision matrix couples basis orders in a spatially
local way, and balances spatial flux and classical electromagnetic terms in
the equation governing the time rate of change of the spectral modes.
The formulation is restricted to the non-relativistic, fast-particle kernel,
consistent with the Bethe-Born approximation \citep{bethe1930}, and uses optical oscillator
strength data as in photon-matter opacity calculations. 
This development has been done in the MASS-APP spectral solver framework,
which uses Kokkos~\citep{trott2021} for thread/GPU parallelism and FleCSI
\citep{bergen2016} to support MPI or Legion-parallel task backends, each layer
providing portability and performance on different HPC systems.
The code has a full treatment of classical $\vec{E}$ and $\vec{B}$ fields, thus
enabling efficient 3D calculations of KN ejecta $\beta$-particle thermalization
along with electromagnetic effects (which we did not explore with the kernel in this work).

We expand the cross section in terms of the Hermite basis, and find that the simple
leading-order dependence of the cross section on atomic bound-bound level transition
energy propagates to the expansion coefficients.
Moreover, the symmetry of the kernel in radial velocity (in velocity space) imply
these expansion coefficients are symmetric under permutation of the mode/order indices,
and hence form a compressible low-cost data structure for computer memory when fitted
to linear functions in log-coefficient-log-transition energy space.
Similarly, the compact triple Hermite product functions, which build the elements of
the spectral scattering matrix, are symmetric under permutation of the mode/order indices,
and obey mode/order parity-based index selection rules that make them sparse.
These compact triple Hermite product functions also permit interoperability of the spectral
collision matrix with adaptive basis coefficients, though we do not test this here.

Numerical results indicate that a reasonable choice of Hermite basis parameters for
$\beta$-particles in the KN are a bulk velocity parameter $\vec{u}=0$, a thermal velocity
parameter $\vec{\alpha}=0.5c$, and a 9x9x9 mode velocity basis set (Hermite orders 0 to 8
in each dimension).
Section~\ref{sec:numver} verifies each step in the computation of the spectral collision
matrix by comparing compact Hermite triple product functions and matrix elements to the
equivalent direct numerical quadrature of the corresponding integrals.
Furthermore, in Section~\ref{sec:numver} we demonstrate the ability to fit the coefficients
of the fast-particle cross section Hermite expansion with linear functions, a property
inherited from the leading-order behavior of the differential cross section for small
ratios of excitation energy to $\beta$-particle kinetic energy.
With an implementation of the spectral collision matrix in MASS-APP, we show a proof of
principle calculation of $\beta$-particle propagation and excitation interaction in a
3D snapshot simulation of KN ejecta.
Given a lower bound scattering angle of $\sim2.5^{\circ}$ and the caveat of the fast-particle
approximation (for instance, using the limit of optical oscillator strengths), this
calculation suggests that large-angle scatters of $\beta$-particles may \textbftog{not be more
than three orders of magnitude lower as a power source for the KN luminosity and spectra,
not including ionization and free electron Coulomb contributions.}

With some further work, the framework should extend to generalized oscillator
strengths and the relativistic kernel given by~\cite{inokuti1971}, with the
caveat that it may be important to replace the Hermite basis with a
causally restricted basis (bounding velocity by the speed of light), for instance
using Legendre polynomials as done by~\cite{manzini2017}.
Another consideration for velocity space is the coordinate system over which the basis
functions are expressed.
The uncollided $\beta$-particle distribution and collision cross section are spherically
symmetric in velocity space (though technically, the collision cross section is symmetric
about the origin of the center-of-mass frame for each collision), so basis functions in
spherical velocity coordinates (for instance, spherical harmonics) may further reduce
the size of the basis expansion required for accurate solutions.

Another improvement to fidelity would be to use the homologous expansion velocity of the
KN ejecta as the value of the bulk velocity parameter $\vec{u}$ in the basis functions, but this
may be generally of negligible effect~\citep{barnes2016}.
Since $\vec{u}$ is variable in space-time, this would require a modification to the standard
spectral system to account for the variation in the equations, but varying the thermal ($\vec{\alpha}$)
and bulk velocity parameters is an active topic of research~\citep{bessemoulin2023,pagliantini2023}.
This would capture some effect of the expansion velocity in the $\beta$-particle simulation, permitting
a Galilean frame transformation of the kernel. \todo{Why bother (maybe rephrase)?}

After adding Coulomb and ionization interactions to the spectral collision matrix,
and incorporating both bulk ejecta and microphysical relativistic effects in the kernel,
we ultimately intend to use MASS-APP to perform detailed thermalization calculations
for $\beta$-particles in KN ejecta, subject to different $\vec{E}$ and $\vec{B}$
field seeds, and use the results to power photon transfer simulations that synthesize
observable light curves and spectra, following~\cite{barnes2021}.

Finally, the efficient representation of the coefficients of the Hermite expansion of
the cross section as linear in log-log space with respect to atomic transition energy
suggests that other cross section formulae may conform at least approximately to an
efficient representation over quantum atomic parameters.

\begin{acknowledgments}
This work was supported by the U.S. Department of Energy through the Los Alamos National Laboratory. Los Alamos National Laboratory is operated by Triad National Security, LLC, for the National Nuclear Security Administration of the U.S. Department of Energy (Contract No. 89233218CNA000001).
Research presented in this article was supported by the Laboratory Directed Research and Development program of Los Alamos National Laboratory under project number 20220104DR.
This research used resources provided by the Los Alamos National Laboratory Institutional Computing Program, which is supported by the U.S. Department of Energy National Nuclear Security Administration under Contract No. 89233218CNA000001.
\end{acknowledgments}

\vspace{5mm}

\software{FleCSI \citep{bergen2016},  
          Kokkos \citep{trott2021}, 
          MASS-APP
          }



\appendix

\section{Kernel integral evaluation and approximation} \label{app:A}

\subsection{Maxwellian integral evaluation}

For completeness, we give details for evaluating the integral over atom/ion velocity here.
We do not assume $\vec{v}_{sa}=\vec{v}_s$, but instead show how the same result can be
obtained after evaluating the integral over $\vec{v}_a$.
Assuming the pre-/post-collision atom/ion distribution is Maxwellian, and
that the comoving collision integral is independent of ion/atom velocity, the
ion/atom-dependent velocity integral is
\begin{equation}
    \label{eq6:awhb}
    \int|\vec{v}_a-\vec{v}_s|e^{-M_av_a^2/2k_BT}d^3\vec{v}_a
    = 2\pi\int_0^{\infty}\int_{-1}^1
    v_{sa}^3e^{-M_a(v_{sa}^2+2v_{sa}v_s\eta+v_s^2)/2k_BT}d\eta dv_{sa} \;\;,
\end{equation}
where $\eta$ is the cosine of the angle between vectors $\vec{v}_{sa}$ and $\vec{v}_s$,
and the integral on the right side is over spherical coordinates.
Integrating first over $\eta$,
\begin{multline}
    \label{eq7:awhb}
    2\pi\int_0^{\infty}\int_{-1}^1
    v_{sa}^3e^{-M_a(v_{sa}^2-2v_{sa}v_s\eta+v_s^2)/2k_BT}d\eta dv_{sa} 
    = 2\pi\int_0^{\infty}
    v_{sa}^3e^{-M_a(v_{sa}^2+v_s^2)/2k_BT}
    \left(\int_{-1}^1e^{M_av_{sa}v_s\eta/k_BT}d\eta\right)dv_{sa} \\
    = 2\pi\int_0^{\infty}v_{sa}^3e^{-M_a(v_{sa}^2+v_s^2)/2k_BT}
    \left(\frac{2k_BT}{M_av_{sa}v_s}\sinh\left(\frac{M_av_{sa}v_s}{k_BT}\right)\right)dv_{sa} \\
    = \frac{4\pi k_BT}{M_av_s}\int_0^{\infty}v_{sa}^2e^{-M_a(v_{sa}^2+v_s^2)/2k_BT}
    \sinh\left(\frac{M_av_{sa}v_s}{k_BT}\right)dv_{sa}
    \;\;.
\end{multline}
Expressing the hyperbolic sine in terms of exponentials, and completing the squares in
the exponents,
\begin{multline}
    \label{eq8:awhb}
    \frac{4\pi k_BT}{M_av_s}\int_0^{\infty}v_{sa}^2e^{-M_a(v_{sa}^2+v_s^2)/2k_BT}
    \frac{1}{2}\left(e^{M_av_{sa}v_s/k_BT}-e^{-M_av_{sa}v_s/k_BT}\right)dv_{sa} \\
    = \frac{2\pi k_BT}{M_av_s}\int_0^{\infty}
    \left(v_{sa}^2e^{-M_a(v_{sa}-v_s)^2/2k_BT}
    -v_{sa}^2e^{-M_a(v_{sa}+v_s)^2/2k_BT}\right)dv_{sa} \;\;,
\end{multline}
where the rightmost integral can be re-expressed as
\begin{multline}
    \label{eq9:awhb}
    \int_0^{\infty}
    \left(v_{sa}^2e^{-M_a(v_{sa}-v_s)^2/2k_BT}
    -v_{sa}^2e^{-M_a(v_{sa}+v_s)^2/2k_BT}\right)dv_{sa} \\
    = \int_{-v_s}^{\infty}(v+v_s)^2e^{-M_av^2/2k_BT}dv
    -\int_{v_s}^{\infty}(v-v_s)^2e^{-M_av^2/2k_BT}dv \\
    = \int_{-v_s}^{v_s}v^2e^{-M_av^2/2k_BT}dv
    + v_s^2\int_{-v_s}^{v_s}e^{-M_av^2/2k_BT}dv
    + 2v_s\left(\int_{-v_s}^{\infty}ve^{-M_av^2/2k_BT}dv
    + \int_{v_s}^{\infty}ve^{-M_av^2/2k_BT}dv\right)
    \;\;,
\end{multline}
using the substitution $v=v_{sa}-v_s$ ($v=v_{sa}+v_s$) in the integral over the first
(second) term.
Using
\begin{subequations}
    \label{eq10:awhb}
    \begin{gather}
        \int_{-v_s}^{v_s}e^{-Cv^2}dv = \sqrt{\frac{\pi}{C}}\erf(\sqrt{C}v_s) \;\;, \\
        \int_{-v_s}^{v_s}ve^{-Cv^2}dv = 0 \;\;, \\
        2\int_{v_s}^{\infty}ve^{-Cv^2}dv = \frac{1}{C}e^{-Cv_s^2} \;\;, \\
        \int_{-v_s}^{v_s}v^2e^{-Cv^2}dv = \frac{1}{2C}\sqrt{\frac{\pi}{C}}\erf(\sqrt{C}v_s)
        - \frac{v_s}{C}e^{-Cv_s^2} \;\;,
    \end{gather}
\end{subequations}
Eq.~\eqref{eq9:awhb} further reduces to
\begin{multline}
    \label{eq11:awhb}
    \int_0^{\infty}
    \left(v_{sa}^2e^{-M_a(v_{sa}-v_s)^2/2k_BT}
    -v_{sa}^2e^{-M_a(v_{sa}+v_s)^2/2k_BT}\right)dv_{sa} \\
    = \int_{-v_s}^{v_s}v^2e^{-M_av^2/2k_BT}dv
    + v_s^2\int_{-v_s}^{v_s}e^{-M_av^2/2k_BT}dv
    + 2v_s\left(\int_{-v_s}^{\infty}ve^{-M_av^2/2k_BT}dv
    + \int_{v_s}^{\infty}ve^{-M_av^2/2k_BT}dv\right) \\
    = \frac{k_BT}{M_a}\sqrt{\frac{2\pi k_BT}{M_a}}\erf\left(\sqrt{\frac{M_a}{2k_BT}}v_s\right)
    - \frac{2k_BTv_s}{M_a}e^{-M_av_s^2/2k_BT}
    + v_s^2\sqrt{\frac{2\pi k_BT}{M_a}}\erf\left(\sqrt{\frac{M_a}{2k_BT}}v_s\right)
    + 2v_s\frac{2k_BT}{M_a}e^{-M_av_s^2/2k_BT} \\
    = \sqrt{\frac{2\pi k_BT}{M_a}}\left(\frac{k_BT}{M_a} + v_s^2\right)
    \erf\left(\sqrt{\frac{M_a}{2k_BT}}v_s\right)
    + \frac{2k_BTv_s}{M_a}e^{-M_av_s^2/2k_BT}
    \;\;.
\end{multline}
Incorporating Eq.~\eqref{eq11:awhb} into Eq.~\eqref{eq6:awhb} (via 
Eqs.~\eqref{eq7:awhb} and \eqref{eq8:awhb}),
\begin{multline}
    \label{eq12:awhb}
    \int|\vec{v}_a-\vec{v}_s|e^{-M_av_a^2/2k_BT}d^3\vec{v}_a \\
    = \frac{2\pi k_BT}{M_av_s}
    \left(\sqrt{\frac{2\pi k_BT}{M_a}}\left(\frac{k_BT}{M_a} + v_s^2\right)
    \erf\left(\sqrt{\frac{M_a}{2k_BT}}v_s\right)
    + \frac{2k_BTv_s}{M_a}e^{-M_av_s^2/2k_BT}\right) \\
    = \left(\frac{2\pi k_BT}{M_a}\right)^{3/2}\frac{1}{v_s}
    \left(\frac{k_BT}{M_a} + v_s^2\right)\erf\left(\sqrt{\frac{M_a}{2k_BT}}v_s\right)
    + 4\pi\left(\frac{k_BT}{M_a}\right)^2e^{-M_av_s^2/2k_BT} \\
    = \left(\frac{2\pi k_BT}{M_a}\right)^{3/2}\overline{\Lambda}_a(v_s, T)
    \;\;,
\end{multline}
where we have introduced the function $\overline{\Lambda}_a(v_s, T)$, which
represents the atom/ion  distribution-weighted average magnitude of difference
in velocity between the $\beta$-particle and the atom/ions.
When $v_s\gg\sqrt{k_BT/M_a}$, Eq.~\eqref{eq12:awhb} simplifies to
\begin{equation}
    \label{eq13:awhb}
    \overline{\Lambda}_a(v_s, T) 
    \approx \frac{1}{v_s}\left(\frac{k_BT}{M_a} + v_s^2\right)
    \approx v_s \;\;.
\end{equation}
Thus, we have obtained the inverse of the Maxwellian normalization factor times
the magnitude of the $\beta$-particle velocity, as desired.

\subsection{Solid angle integral of cross section}

In Section~\ref{sec:method}, we factor the differential cross section out of
the integral over atom/ion velocity, which can be evaluated as
\begin{equation}
    \label{eq18:awhb}
    \int\frac{d\sigma_{jj'}}{d\Omega}d\Omega
    = 4\pi f_{jj'} \left(\frac{e^2}{4E_k}\right)^2\left(\frac{E_k}{E_{jj'}}\right)
    \int_{\theta_{\min}}^{\pi}\csc^4(\theta/2)
    \left[1-\frac{1}{2}\left(\frac{E_{jj'}}{E_k}\right)
    -\left(\sqrt{1-\left(\frac{E_{jj'}}{E_k}\right)}\right)\cos(\theta)
    \right]\sin(\theta)d\theta \;\;,
\end{equation}
where $\theta_{\min}$ is a minimum scattering angle, which in general may
depend on $E_{jj'}$ (but we set it to a constant in this work).
Making use of
\begin{subequations}
    \label{eq19:awhb}
    \begin{gather}
        \int_{\theta_{\min}}^{\pi}\csc^4(\theta/2)\sin(\theta)d\theta 
        = 4\int_{-1}^{1-\varepsilon}\frac{d\mu}{(1-\mu)^2}
        = 4\left(\frac{1}{\varepsilon} - \frac{1}{2}\right) \;\;, \\
        \int_{\theta_{\min}}^{\pi}\csc^4(\theta/2)\cos(\theta)\sin(\theta)d\theta
        = 4\int_{-1}^{1-\varepsilon}\frac{\mu d\mu}{(1-\mu)^2}
        = 4\left(\frac{1-\varepsilon}{\varepsilon} + \frac{1}{2} + \ln(\varepsilon/2)\right)
        \;\;,
    \end{gather}
\end{subequations}
where $\mu = \cos(\theta)$ and $1-\varepsilon = \cos(\theta_{\min})$,
Eq.~\eqref{eq18:awhb} becomes
\begin{multline}
    \label{eq20:awhb}
    \int\frac{d\sigma_{jj'}}{d\Omega}d\Omega
    = \sigma_{jj'}(v_s) = \\
    8\pi f_{jj'} \left(\frac{e^2}{2mv_s^2}\right)^2
    \left\{
    \left[\left(\frac{mv_s^2}{E_{jj'}}\right) - 1\right]
    \left(\frac{1}{\varepsilon} - \frac{1}{2}\right)
    - \left[\left(\frac{mv_s^2}{E_{jj'}}\right)
    \sqrt{1-\left(\frac{2E_{jj'}}{mv_s^2}\right)}\right]
    \left(\frac{1-\varepsilon}{\varepsilon} + \frac{1}{2} + \ln(\varepsilon/2)\right)
    \right\} \;\;,
\end{multline}
where we have used $E_k = m|\vec{v}_s-\overline{\vec{v}}_a|^2/2 = mv_s^2/2$.

\section{Derivation of compact Hermite triple products} \label{app:B}

\subsection{Kernel expansion and standard triple-Hermite product}

Expanding the pre-scatter kernel function, written with explicit dependence on a transition
energy, in the upper-index basis,
\begin{equation}
    \label{eq1:appB}
    S_c^{\rm (pre)}(v_s, E_{jj'}) \approx \sum_{i_x'', i_y'', i_z''}^{I_x, I_y, I_z}
    \Psi^{i_x'', i_y'', i_z''}(\vec{\xi}_s)\mathcal{D}_{i_x'', i_y'', i_z''}(E_{jj'}) \;\;,
\end{equation}
where $\mathcal{D}_{i_x'', i_y'', i_z''}(E_{jj'})$ depends on transition energy but
not on $\beta$-particle speed $v_s$.
Incorporating Eq.~\eqref{eq1:appB} into the integral for the spectral matrix elements,
\begin{multline}
    \label{eq2:appB}
    S_{i_x',i_y',i_z'}^{i_x,i_y,i_z \rm (pre)}
    = \int\Psi^{i_x,i_y,i_z}(\vec{\xi}_s)S_c^{\rm (pre)}(v_s, E_{jj'})
    \Psi_{i_x',i_y',i_z'}(\vec{\xi}_s)d^3\vec{\xi}_s \\
    = \sum_{i_x'', i_y'', i_z''}^{I_x, I_y, I_z}\mathcal{D}_{i_x'', i_y'', i_z''}(E_{jj'})
    \int\Psi^{i_x,i_y,i_z}(\vec{\xi}_s)\Psi^{i_x'', i_y'', i_z''}(\vec{\xi}_s)
    \Psi_{i_x',i_y',i_z'}(\vec{\xi}_s)d^3\vec{\xi}_s \;\;.
\end{multline}
Using the standard formulae,
\begin{subequations}
    \label{eq3:appB}
    \begin{gather}
        \psi^{i_q}(\xi_q) = \frac{1}{\sqrt{2^{i_q}i_q!}}H_{i_q}(\xi_q) \;\;, \\
        \psi_{i_q}(\xi_q) = \frac{1}{\sqrt{\pi 2^{i_q}i_q!}}H_{i_q}(\xi_q)e^{-\xi_q^2} 
        = \frac{1}{\sqrt{\pi}}\psi^{i_q}(\xi_q)e^{-\xi_q^2} \;\;, \\
        \Psi^{i_x,i_y,i_z}(\vec{\xi}_s) = \psi^{i_x}(\xi_x)\psi^{i_y}(\xi_y)\psi^{i_z}(\xi_z) \;\;, \\
        \Psi_{i_x,i_y,i_z}(\vec{\xi}_s) = \psi_{i_x}(\xi_x)\psi_{i_y}(\xi_y)\psi_{i_z}(\xi_z) \;\;,
    \end{gather}
\end{subequations}
where $q$ denotes $x$, $y$, or $z$, $H_{i_q}$ is the Hermite polynomial of order $i_q$, and
$\vec{\xi}_s = (\xi_x, \xi_y, \xi_z)$.
Incorporating Eqs.~\eqref{eq3:appB} into the right side of Eq.~\eqref{eq2:appB},
\begin{multline}
    \label{eq4:appB}
    S_{i_x',i_y',i_z'}^{i_x,i_y,i_z \rm (pre)}
    = \sum_{i_x'', i_y'', i_z''}^{I_x, I_y, I_z}\mathcal{D}_{i_x'', i_y'', i_z''}(E_{jj'})
    \int\Psi^{i_x,i_y,i_z}(\vec{\xi}_s)\Psi^{i_x'', i_y'', i_z''}(\vec{\xi}_s)
    \Psi_{i_x',i_y',i_z'}(\vec{\xi}_s)d^3\vec{\xi}_s \\
    = \sum_{i_x'', i_y'', i_z''}^{I_x, I_y, I_z}\mathcal{D}_{i_x'', i_y'', i_z''}(E_{jj'})
    \int\int\int\psi^{i_x}(\xi_x)\psi^{i_y}(\xi_y)\psi^{i_z}(\xi_z)
    \psi^{i_x''}(\xi_x)\psi^{i_y''}(\xi_y)\psi^{i_z''}(\xi_z)
    \psi_{i_x'}(\xi_x)\psi_{i_y'}(\xi_y)\psi_{i_z'}(\xi_z) d\xi_x d\xi_y d\xi_z \\
    = \sum_{i_x'', i_y'', i_z''}^{I_x, I_y, I_z}\mathcal{D}_{i_x'', i_y'', i_z''}(E_{jj'})
    \left(\int\psi^{i_x}(\xi_x)\psi^{i_x''}(\xi_x)\psi_{i_x'}(\xi_x)d\xi_x\right)
    \left(\int\psi^{i_y}(\xi_y)\psi^{i_y''}(\xi_y)\psi_{i_y'}(\xi_y)d\xi_y\right)
    \left(\int\psi^{i_z}(\xi_z)\psi^{i_z''}(\xi_z)\psi_{i_z'}(\xi_z)d\xi_z\right) \;\;,
\end{multline}
where each 1V integral can be expressed in terms of Hermite polynomials as
\begin{equation}
    \label{eq5:appB}
    \int\psi^{i_q}(\xi_q)\psi^{i_q''}(\xi_q)\psi_{i_q'}(\xi_q)d\xi_q
    = \frac{1}{\sqrt{2^{i_q}i_q!}\sqrt{2^{i_q''}i_q''!}\sqrt{\pi 2^{i_q'}i_q'!}}
    \int H_{i_q}(\xi_q)H_{i_q''}(\xi_q)H_{i_q'}(\xi_q)e^{-\xi_q^2}d\xi_q \;\;.
\end{equation}
The triple-Hermite integral on the right side can be simplified using the result of
\cite{askey1999} (Chapter 6),
\begin{multline}
    \label{eq6:appB}
    \int H_{i_q}(\xi_q)H_{i_q''}(\xi_q)H_{i_q'}(\xi_q)e^{-\xi_q^2}d\xi_q \\
    = 
    \begin{cases}
        \displaystyle
        \frac{2^{(i_q+i_q'+i_q'')/2}i_q!i_q'!i_q''!\sqrt{\pi}}{((-i_q+i_q'+i_q'')/2)!((i_q-i_q'+i_q'')/2)!((i_q+i_q'-i_q'')/2)!} \;\;, \\
        \;\;\;\; \text{if } i_q+i_q'+i_q'' \text{ is even and }
        i_a+i_b\geq i_c \;\;\forall\; (i_a,i_b,i_c)\in\{\text{permutations of} (i_q,i_q',i_q'')\}\;\;,
        \\\\
        0 \;\;,\;\; \text{otherwise.}
    \end{cases}
\end{multline}
This formula for the triple-Hermite integral is symmetric under permutations of $(i_q,i_q',i_q'')$,
and the requirement of the denominator factorial arguments being positive is equivalent to a discrete
triangle inequality for ``side lengths'' $i_q$, $i_q'$, and $i_q''$.
Moreover, it follows from basic parity arguments that 
$i_q+i_q'+i_q'' \equiv 0 \mod 2 \rightarrow \pm i_q \pm i_q' \pm i_q''  \equiv 0 \mod 2$ (replacing
pluses with minuses preserves evenness).
Incorporating Eq.~\eqref{eq6:appB} into Eq.~\eqref{eq5:appB}, the powers of 2 and $\pi$ cancel,
\begin{multline}
    \label{eq7:appB}
    T(i_q,i_q',i_q'') \equiv \int\psi^{i_q}(\xi_q)\psi^{i_q''}(\xi_q)\psi_{i_q'}(\xi_q)d\xi_q \\
     =
    \begin{cases}
        \displaystyle
        \frac{\sqrt{i_q!i_q'!i_q''!}}{((-i_q+i_q'+i_q'')/2)!((i_q-i_q'+i_q'')/2)!((i_q+i_q'-i_q'')/2)!} \;\;, \\
        \;\;\;\; \text{if } i_q+i_q'+i_q'' \text{ is even and }
        i_a+i_b\geq i_c \;\;\forall\; (i_a,i_b,i_c)\in\{\text{permutations of} (i_q,i_q',i_q'')\}\;\;,
        \\\\
        0 \;\;,\;\; \text{otherwise,}
    \end{cases}
\end{multline}
where $T(i_q,i'_q,i_q'')$ is symmetric over all permutations of $(i_q,i_q',i_q'')$.
Using the newly defined $T(i_q,i'_q,i_q'')$ in Eq.~\eqref{eq4:appB}
\begin{equation}
    \label{eq8:appB}
    S_{i_x',i_y',i_z'}^{i_x,i_y,i_z \rm (pre)}
    = \sum_{i_x'', i_y'', i_z''}^{I_x, I_y, I_z}\mathcal{D}_{i_x'', i_y'', i_z''}(E_{nn'})
    T(i_x,i_x',i_x'')T(i_y,i_y',i_y'')T(i_z,i_z',i_z'') \;\;.
\end{equation}

\subsection{Kernel expansion with compact triple-Hermite product}

Expanding the pre-scatter kernel function in the lower-index basis,
\begin{equation}
    \label{eq9:appB}
    S_c^{\rm (pre)}(v_s, E_{jj'}) \approx \sum_{i_x'', i_y'', i_z''}^{I_x, I_y, I_z}
    \mathcal{D}_{i_x'', i_y'', i_z''}(E_{jj'})\Psi_{i_x'', i_y'', i_z''}(\vec{\xi}_s) \;\;,
\end{equation}
where, as in the preceding section, the $\mathcal{D}_{i_x'', i_y'', i_z''}(E_{jj'})$ coefficients
depend on transition energy but not on $\beta$-particle speed $v_s$.
The steps for expanding the pre-scatter matrix follow the upper-index formulation, but each 1V
integral is now
\begin{equation}
    \label{eq10:appB}
    \int\psi^{i_q}(\xi_q)\psi_{i_q''}(\xi_q)\psi_{i_q'}(\xi_q)d\xi_q
    = \frac{1}{\sqrt{2^{i_q}i_q!}\sqrt{\pi 2^{i_q''}i_q''!}\sqrt{\pi 2^{i_q'}i_q'!}}
    \int H_{i_q}(\xi_q)H_{i_q''}(\xi_q)H_{i_q'}(\xi_q)e^{-2\xi_q^2}d\xi_q \;\;.
\end{equation}
The factor of 2 in the exponent of the integral weight implies Eq.~\eqref{eq6:appB} cannot be
applied directly to Eq.~\eqref{eq10:appB}.
Following~\cite{askey1999}, we may use a 3-variable generator function to evaluate the integral
on the right side of Eq.~\eqref{eq10:appB} as coefficients of the expansion of the generator function,
\begin{equation}
    \label{eq11:appB}
    F(r,s,t) = \sum_{i_q=0}^{\infty}\sum_{i_q'=0}^{\infty}\sum_{i_q''=0}^{\infty}
    \frac{1}{i_q!i_q'!i_q''!}
    \left(\int H_{i_q}(\xi_q)H_{i_q''}(\xi_q)H_{i_q'}(\xi_q)e^{-2\xi_q^2}d\xi_q\right)
    r^{i_q}s^{i_q'}t^{i_q''} \;\;,
\end{equation}
where $r$, $s$, and $t$ are the formal variables.
Interchanging the sums gives
\begin{equation}
    \label{eq12:appB}
    F(r,s,t) = \int
    \left(\sum_{i_q=0}^{\infty}\frac{1}{i_q!}H_{i_q}(\xi_q)r^{i_q}
    \sum_{i_q'=0}^{\infty}\frac{1}{i_q'!}H_{i_q'}(\xi_q)s^{i_q'}
    \sum_{i_q''=0}^{\infty}\frac{1}{i_q''!}H_{i_q''}(\xi_q)t^{i_q''}\right)
    e^{-2\xi_q^2}d\xi_q \;\;.
\end{equation}
From~\cite{askey1999}, the sums in parentheses can be evaluated as exponentials,
\begin{equation}
    \label{eq13:appB}
    F(r,s,t) = \int
    \left(e^{2r\xi_q-r^2}e^{2s\xi_q-s^2}e^{2t\xi_q-t^2}\right)e^{-2\xi_q^2}d\xi_q \;\;.
\end{equation}
We complete the square in a different way than~\cite{askey1999}, and evaluate the integral as follows,
\begin{multline}
    \label{eq14:appB}
    F(r,s,t) = \int e^{-2\xi_q^2 + 2r\xi_q-r^2 + 2s\xi_q-s^2 + 2t\xi_q-t^2}d\xi_q
    = \int e^{-(4\xi_q^2 - 4r\xi_q - 4s\xi_q - 4t\xi_q)/2 - r^2 - s^2 - t^2}d\xi_q \\
    = \int e^{-(4\xi_q^2 - 4(r+s+t)\xi_q + (r+s+t)^2)/2 + (r+s+t)^2/2 - (r^2+s^2+t^2)}d\xi_q
    = e^{(r+s+t)^2/2 - (r^2+s^2+t^2)}\int e^{-(2\xi_q - (r+s+t))^2/2}d\xi_q \\
    = e^{rs+st+tr - (r^2+s^2+t^2)/2}\int e^{-2(\xi_q - (r+s+t)/2)^2}d\xi_q 
    = e^{rs+st+tr - (r^2+s^2+t^2)/2}\sqrt{\frac{\pi}{2}} \;\;.
\end{multline}
In~\cite{askey1999}, only the cross-multiplication terms remained in their equivalent of
Eq.~\eqref{eq14:appB}, which enabled an expansion of the exponential as a product of three series,
hence three series indices that could be related to $(i_q,i_q',i_q'')$.
Here, we introduce six indices: three for the cross terms and three for the diagonal terms, which
makes the system of equations relating $(i_q,i_q',i_q'')$ underdetermined (unlike~\cite{askey1999}).
The resulting form of the generator function is
\begin{multline}
    \label{eq15:appB}
    F(r,s,t) = \sqrt{\frac{\pi}{2}}e^{rs+st+tr - (r^2+s^2+t^2)/2} \\
    = \sqrt{\frac{\pi}{2}}\sum_{a=0}^{\infty}\frac{(rs)^a}{a!}\sum_{b=0}^{\infty}\frac{(st)^b}{b!}
    \sum_{c=0}^{\infty}\frac{(tr)^c}{c!}\sum_{b'=0}^{\infty}\frac{(-1/2)^{b'}r^{2b'}}{b'!}
    \sum_{\tilde{c'}=0}^{\infty}\frac{(-1/2)^{c'}s^{2c'}}{c'!}
    \sum_{a'=0}^{\infty}\frac{(-1/2)^{a'}t^{2a'}}{a'!} \\
    = \sum_{a=0}^{\infty}\sum_{b=0}^{\infty}\sum_{c=0}^{\infty}
    \sum_{a'=0}^{\infty}\sum_{b'=0}^{\infty}\sum_{c'=0}^{\infty}\sqrt{\frac{\pi}{2}}
    \left(-\frac{1}{2}\right)^{a'+b'+c'}\frac{r^{c+a+2b'}s^{a+b+2c'}t^{b+c+2a'}}{a!b!c!a'!b'!c'!}
    \;\;.
\end{multline}
We now introduce the following system of equations in order to re-index the sum,
\begin{subequations}
    \label{eq16:appB}
    \begin{gather}
        i_q = c + a + 2b' \;\;, \\
        i_q' = a + b + 2c' \;\;, \\
        i_q'' = b + c + 2a' \;\;.
    \end{gather}
\end{subequations}
Solving Eqs.~\eqref{eq16:appB} for $(a,b,c)$ in terms of $(i_q,i_q',i_q'',a',b',c')$,
\begin{subequations}
    \label{eq17:appB}
    \begin{gather}
        a = \frac{i_q + i_q' - i_q''}{2} - (b' + c' - a') \;\;, \\
        b = \frac{i_q' + i_q'' - i_q}{2} - (a' + c' - b') \;\;, \\
        c = \frac{i_q + i_q'' - i_q'}{2} - (a' + b' - c') \;\;.
    \end{gather}
\end{subequations}
From Eqs.~\eqref{eq17:appB}, we obtain the constraint that $i_q+i_q'+i_q''$ must be
even, as in the upper-index formulation~\cite{askey1999}.
Furthermore, we notice from Eqs.~\eqref{eq16:appB} that $2b'\leq i_q$, $2c'\leq i_q'$, and
$2a'\leq i_q''$.
These index relations imply we can re-index the sum as follows,
\begin{multline}
    \label{eq18:appB}
    F(r,s,t)
    = \sum_{i_q=0}^{\infty}\sum_{i_q'=0}^{\infty}\sum_{iq''=0}^{\infty}
    \delta_{i_q+i_q'+i_q'' \,,\, 2\lfloor(i_q+i_q'+i_q'')/2\rfloor}\sqrt{\frac{\pi}{2}}\left(
    \sum_{a'=0}^{\lfloor i_q''/2\rfloor}\sum_{b'=0}^{\lfloor i_q/2\rfloor}
    \sum_{c'=0}^{\lfloor i_q'/2\rfloor}
    \left(-\frac{1}{2}\right)^{a'+b'+c'}\frac{1}{a'!b'!c'!} \right.\\\left.
    \frac{\Theta(i_q+i_q'-i_q''-2(b'+c'-a'))\,\Theta(i_q'+i_q''-i_q-2(a'+c'-b'))
    \,\Theta(i_q+i_q''-i_q'-2(a'+b'-c'))}
    {((i_q+i_q'-i_q'')/2-(b'+c'-a'))!\,((i_q'+i_q''-i_q)/2-(a'+c'-b'))!
    \,((i_q+i_q''-i_q')/2-(a'+b'-c'))!}\right)
    r^{i_q}s^{i_q'}t^{i_q''}
    \;\;,
\end{multline}
where $\delta_{i_q+i_q'+i_q'', 2\lfloor(i_q+i_q'+i_q'')/2\rfloor}$ is the Kronecker delta
function that is 0 (1) when $i_q+i_q'+i_q''$ is odd (even) and $\Theta(\cdot)$ is the
discrete step function, which is 0 (1) for negative (positive) argument.
We may now unambiguously relate the modified triple integral to analytic closed form coefficients,
\begin{multline}
    \label{eq19:appB}
    \int H_{i_q}(\xi_q)H_{i_q''}(\xi_q)H_{i_q'}(\xi_q)e^{-2\xi_q^2}d\xi_q = \\ 
    i_q!i_q'!i_q''! \,
    \delta_{i_q+i_q'+i_q'' \,,\, 2\lfloor(i_q+i_q'+i_q'')/2\rfloor}\sqrt{\frac{\pi}{2}}\left(
    \sum_{a'=0}^{\lfloor i_q''/2\rfloor}\sum_{b'=0}^{\lfloor i_q/2\rfloor}
    \sum_{c'=0}^{\lfloor i_q'/2\rfloor}
    \left(-\frac{1}{2}\right)^{a'+b'+c'}\frac{1}{a'!b'!c'!} \right.\\\left.
    \frac{\Theta(i_q+i_q'-i_q''-2(b'+c'-a'))\,\Theta(i_q'+i_q''-i_q-2(a'+c'-b'))
    \,\Theta(i_q+i_q''-i_q'-2(a'+b'-c'))}
    {((i_q+i_q'-i_q'')/2-(b'+c'-a'))!\,((i_q'+i_q''-i_q)/2-(a'+c'-b'))!
    \,((i_q+i_q''-i_q')/2-(a'+b'-c'))!}\right) \;\;.
\end{multline}
The discrete step functions encode the discrete triangle inequality condition discussed in the
previous section, but with $(i_q,i_q',i_q'')$ replaced with $(i_q-2b',i_q'-2c',i_q''-2a')$.
Using Eq.~\eqref{eq19:appB}, Eq.~\eqref{eq10:appB} can be written as
\begin{multline}
    \label{eq20:appB}
    \int\psi^{i_q}(\xi_q)\psi_{i_q''}(\xi_q)\psi_{i_q'}(\xi_q)d\xi_q = \\
    \frac{1}{\sqrt{2\pi}2^{(i_q+i_q'+i_q'')/2}}
    \sqrt{i_q!i_q'!i_q''!} \,
    \delta_{i_q+i_q'+i_q'' \,,\, 2\lfloor(i_q+i_q'+i_q'')/2\rfloor}\left(
    \sum_{a'=0}^{\lfloor i_q''/2\rfloor}\sum_{b'=0}^{\lfloor i_q/2\rfloor}
    \sum_{c'=0}^{\lfloor i_q'/2\rfloor}
    \left(-\frac{1}{2}\right)^{a'+b'+c'}\frac{1}{a'!b'!c'!} \right.\\\left.
    \frac{\Theta(i_q+i_q'-i_q''-2(b'+c'-a'))\,\Theta(i_q'+i_q''-i_q-2(a'+c'-b'))
    \,\Theta(i_q+i_q''-i_q'-2(a'+b'-c'))}
    {((i_q+i_q'-i_q'')/2-(b'+c'-a'))!\,((i_q'+i_q''-i_q)/2-(a'+c'-b'))!
    \,((i_q+i_q''-i_q')/2-(a'+b'-c'))!}\right) \;\;.
\end{multline}
This argument of the sum can be expressed in terms of the $T(\cdot,\cdot,\cdot)$ function defined
in Eq.~\eqref{eq7:appB} in the preceding section,
\begin{multline}
    \label{eq20:appB}
    T_c(i_q,i_q',i_q'') = \int\psi^{i_q}(\xi_q)\psi_{i_q''}(\xi_q)\psi_{i_q'}(\xi_q)d\xi_q = \\
    \frac{1}{\sqrt{2\pi}2^{(i_q+i_q'+i_q'')/2}}
    \sqrt{i_q!i_q'!i_q''!} \,
    \left(
    \sum_{a'=0}^{\lfloor i_q''/2\rfloor}\sum_{b'=0}^{\lfloor i_q/2\rfloor}
    \sum_{c'=0}^{\lfloor i_q'/2\rfloor}
    \frac{(-1/2)^{a'+b'+c'}T(i_q-2b',i_q'-2c',i_q''-2a')}
    {a'!b'!c'!\,\sqrt{(i_q-2b')!(i_q'-2c')!(i_q''-2a')!}}
    \right) \;\;,
\end{multline}
where the Kronecker delta is included effectively in $T(\cdot,\cdot,\cdot)$, since the condition
for $i_q+i'_q+i_q''$ to be even is the same as $i_q-2b'+i'_q-2c'+i_q''-2a'$ (though it might be
computationally expedient to check the parity of $i_q+i'_q+i_q''$ prior to any other calculation step).
The pre-scatter matrix is obtained by replacing $T(\cdot)$ with $T_c(\cdot)$ in
Eq.~\eqref{eq8:appB}.

\section{Full Spectrally Discrete Equations with $\vec{E}$ and $\vec{B}$} \label{app:C}

The non-dimensonalization of the Maxwell-Boltzmann equations used in MASS-APP is
\todo{Cite article or book with these non-dimensionalizations.}
\begin{subequations}
    \label{eq0:appC}
    \begin{gather}
        \tilde{t} = \omega_{pe}t \;\;, \\
        \tilde{\vec{x}} = \frac{\omega_{pe}}{c}\vec{x} \;\;, \\
        \tilde{\vec{v}} = \frac{1}{c}\vec{v} \;\;, \\
        \tilde{f}_s(\tilde{\vec{x}}, \tilde{\vec{v}}, \tilde{t})
        = \frac{c^3}{N_0}f_s(\vec{x}, \vec{v}, t) \;\;, \\
        \tilde{q}_s = \frac{q_s}{e} \;\;, \\
        \tilde{m}_s = \frac{m_s}{m_e} \;\;,\;\;
        \tilde{M}_a = \frac{M_s}{m_e} \;\;,\;\;
        \tilde{M} = \frac{M}{m_e} \;\;, \\
        \tilde{\vec{E}} = \frac{1}{c}\sqrt{\frac{\varepsilon_0}{m_eN_0}}\vec{E} \;\;,\;\;
        \tilde{\vec{B}} = \sqrt{\frac{\varepsilon_0}{m_eN_0}}\vec{B} \;\;,
    \end{gather}
\end{subequations}
where $\omega_{pe}$ is a reference plasma electron oscillation frequency, $c$ is the speed of
light, $N_0$ is a reference number density, \textbftog{$e$ is electron charge,} $m_e$ is electron mass,
and $\varepsilon_0$ is permittivity of free space.

The full equations solved in MASS-APP are (dropping the tilde for non-dimensionality)
\begin{subequations}
    \label{eq1:appC}
    \begin{gather}
        \frac{\partial C_{n,m,p}}{\partial t}
        + \nabla \cdot \left( 
        \begin{bmatrix}
            \alpha_x \sqrt{\frac{n+1}{2}} C_{n+1,m,p} \\
            \alpha_y \sqrt{\frac{m+1}{2}} C_{n,m+1,p} \\
            \alpha_z \sqrt{\frac{p+1}{2}} C_{n,m,p+1}
        \end{bmatrix}
        + C_{n,m,p}
        \begin{bmatrix}
            u_x \\ u_y \\ u_z 
        \end{bmatrix}
        +
        \begin{bmatrix}
            \alpha_x \sqrt{\frac{n}{2}} C_{n-1,m,p} \\
            \alpha_y \sqrt{\frac{m}{2}} C_{n,m-1,p} \\
            \alpha_z \sqrt{\frac{p}{2}} C_{n,m,p-1}
        \end{bmatrix}
        \right)
        - \frac{q_s}{m_s} \left (\vec{E} + \begin{bmatrix}
        u_x \\
        u_y \\
        u_z 
        \end{bmatrix}\times\vec{B}\right) \cdot 
        \begin{bmatrix}
            \frac{\sqrt{2n}}{\alpha_x} C_{n-1,m,p} \\ 
            \frac{\sqrt{2m}}{\alpha_y} C_{n,m-1,p} \\ 
            \frac{\sqrt{2p}}{\alpha_z} C_{n,m,p-1}
        \end{bmatrix}
        \nonumber\\
        - \frac{q_s}{m_s} 
        \left( 
        \begin{bmatrix}
            \alpha_x \\
            0 \\
            0
        \end{bmatrix}
        \times \vec{B} \cdot 
        \begin{bmatrix}
            0 \\ 
            \left(\sqrt{nm} C_{n-1,m-1,p} + \sqrt{(n+1)m} C_{n+1,m-1,p} \right )/\alpha_y \\ 
            \left(\sqrt{np} C_{n-1,m,p-1} + \sqrt{(n+1)p} C_{n+1,m,p-1} \right )/\alpha_z  
        \end{bmatrix} 
        \right)
        \nonumber\\
        - \frac{q_s}{m_s} 
        \left( 
        \begin{bmatrix}
            0 \\
            \alpha_y \\
            0
        \end{bmatrix}
        \times \vec{B} \cdot 
        \begin{bmatrix}
        \left(\sqrt{nm} C_{n-1,m-1,p} + \sqrt{n(m+1)} C_{n-1,m+1,p} \right)/\alpha_x \\ 
        0 \\ 
        \left (\sqrt{mp} C_{n,m-1,p-1} + \sqrt{(m+1)p} C_{n,m+1,p-1} \right )/\alpha_z
        \end{bmatrix} 
        \right)
        \nonumber\\
        - \frac{q_s}{m_s} 
        \left( 
        \begin{bmatrix}
            0 \\
            0 \\
            \alpha_z 
        \end{bmatrix}
        \times \vec{B} \cdot 
        \begin{bmatrix}
            \left(\sqrt{np} C_{n-1,m,p-1} + \sqrt{n(p+1)} C_{n-1,m,p+1} \right)/\alpha_x \\ 
            \left(\sqrt{mp} C_{n,m-1,p-1} + \sqrt{m(p+1)} C_{n,m-1,p+1} \right)/\alpha_y \\ 
            0
            \end{bmatrix} 
        \right) = \sum_{n',m',p'}S_{n,m,p}^{n',m',p'}C_{n',m',p'} \;\;, \\
        \frac{\partial\vec{B}}{\partial t} = -c\nabla \times \vec{E} \;\;, \\
        \frac{\partial\vec{E}}{\partial t} =  c\nabla \times \vec{B} 
        - 4 \pi q_s \alpha_x  \alpha_y  \alpha_z \left (C_{0,0,0} 
        \begin{bmatrix}
            u_x \\
            u_y \\
            u_z 
        \end{bmatrix}
        + \frac{1}{\sqrt{2}} 
        \begin{bmatrix}
            \alpha_x C_{1,0,0} \\
            \alpha_y C_{0,1,0} \\
            \alpha_z C_{0,0,1} 
        \end{bmatrix}\right) \;\;,
    \end{gather}
\end{subequations}


\todo{Split reference files for easier readability.}
\bibliography{refs}{}
\bibliographystyle{aasjournal}



\end{document}